\documentclass[pra,twocolumn, aps, preprintnumbers,superscriptaddress]{revtex4-1}
\usepackage[colorlinks=true,citecolor=blue,urlcolor=black]{hyperref}
\usepackage{verbatim}
\usepackage{amsmath}
\usepackage{latexsym}
\usepackage{revsymb}
\usepackage{yfonts}
\usepackage{ifthen}

\usepackage{natbib}
\usepackage{amsfonts}
\usepackage{amsmath}
\usepackage{amssymb}
\usepackage{amsthm}
\usepackage{graphicx}
\usepackage{bm}
\usepackage{bbm}
\usepackage{epsfig,color,amssymb}
\usepackage{subfigure}
\usepackage{amsfonts}
\usepackage{amscd}
\usepackage{amsmath}
\usepackage{multirow}
\usepackage{chemarrow}
\usepackage{dcolumn}
\usepackage{bm}
\usepackage{graphicx}
\usepackage{enumerate}
\usepackage{epsfig}
\usepackage{subfigure}
\usepackage{xcolor}
\usepackage{multirow}
\usepackage{ulem}
\usepackage{braket}
\usepackage{comment}
\usepackage{enumitem}
\usepackage{amsthm}
\renewcommand{\emph}[1]{\textit{#1}}

\begin{document}

\title{Secure quantum secret sharing without signal disturbance monitoring}

\author{Jie Gu}
\author{Yuan-Mei Xie}
\author{Wen-Bo Liu}
\author{Yao Fu}
\author{Hua-Lei Yin}\email{hlyin@nju.edu.cn}
\author{Zeng-Bing Chen}\email{zbchen@nju.edu.cn}
\affiliation{National Laboratory of Solid State Microstructures, School of Physics, and Collaborative Innovation Center of Advanced Microstructures, Nanjing University, Nanjing 210093, China}

\begin{abstract}
Quantum secret sharing (QSS) is an essential primitive for the future quantum internet, which promises secure multiparty communication. However, developing a large-scale QSS network is a huge challenge due to the channel loss and the requirement of multiphoton interference or high-fidelity multipartite entanglement distribution. Here, we propose a three-user QSS protocol without monitoring signal disturbance, which is capable of ensuring the unconditional security. The final key rate of our protocol can be demonstrated to break the Pirandola-Laurenza-Ottaviani-Banchi bound of quantum channel and its simulated transmission distance can approach over 600 km using current techniques. Our results pave the way to realizing high-rate and large-scale QSS networks.
\end{abstract}

\maketitle
\section{INTRODUCTION}
Quantum internet holds numerous advantages over the classical internet in distributing, sharing and processing information~\cite{kimble2008quantum,wehner2018quantum}. Quantum internet consists of quantum networks for quantum computing and quantum communication, with quantum computing offering high calculation speeds~\cite{ladd2010quantum,arute2019quantum} and quantum communication providing robust security~\cite{bennett1984quantum,ekert1991quantum,sasaki2014practical,yin2016detector}. In the realm of quantum communication, besides quantum key distribution which has been well developed on the way to practical applications recently~\cite{yin2016measurement,boaron2018secure,frohlich2013quantum,liao2018satellite,yuan201810,zhang2020long}, quantum secret sharing (QSS) is another cryptographic primitive used for multiparty quantum communication in quantum internet.

Secret sharing aims to split a secret message of one user, called dealer, into several parts and distribute them to other users, called players, with one player receiving one part~\cite{shamir1979share,blakley1979safeguarding}. In secret sharing, it is supposed to guarantee that any unauthorized subset of players cannot reconstruct the message. Only a few players cooperate together can they recover the original secret message. Therefore, classical secret sharing plays a fundamental role in modern information society with applications such as secure multiparty computation, blockchain, cloud storage and computing. Compared with classical secret sharing, QSS will further provide unconditional security based on laws of quantum mechanics~\cite{pirandola2020advances}. The first QSS protocol was proposed in 1999 where three participants share classical information using the three-particle Greenberger-Horne-Zeilinger (GHZ) state~\cite{hillery1999quantum}. Then, a simplified QSS protocol with two-particle entangled states~\cite{karlsson1999quantum} was proposed. In the past few decades, QSS has been widely researched in theory~\cite{chen2007multi,markham2008graph,fu2015long,tavakoli2015secret,kogias2017unconditional,grice2019quantum,wu2020passive}.

Despite that QSS has achieved significant advances, only a little bit of proof-of-principle experimental demonstrations have been conducted including GHZ state~\cite{chen2005experimental,gaertner2007experimental}, single-photon state~\cite{schmid2005experimental}, graph state~\cite{bell2014experimental,cai2017multimode} and bound entanglement state~\cite{zhou2018quantum}.There are reasons in many aspects. First, the entangled state, especially multipartite entangled state, is still hard to be generated, manipulated and distributed with high-brightness, high-fidelity and long-distance~\cite{pan2012multiphoton}. Although one can utilize the post-selected entanglement state~\cite{fu2015long} to circumvent the above issues, it still needs valuable multiphoton interference, which only results in low secure key rate. Second, it is not a trivial task to extract secure key with perfect privacy~\cite{fu2015long,kogias2017unconditional}. Besides, QSS is a multiparty protocol that has at least three users and one cannot assume all participants are honest, which is the biggest problem to design a secure protocol. For example, the original three-user QSS protocol based on GHZ state is not secure if one of the player is dishonest~\cite{qin2007cryptanalysis}. In recent years, the scheme with sequential transmission of single qudit~\cite{schmid2005experimental,tavakoli2015secret,wei2018quantum} have tried to remove the GHZ state requirement. However, it is shown that security of single-qudit schemes have drawbacks and remains to further debate~\cite{he2007comment,schmid2007schmid}. More seriously, single-qudit schemes are vulnerable to Trojan horse attacks~\cite{xu2020secure}.

Twin-field quantum key distribution~\cite{lucamarini2018overcoming} and its variants~\cite{wang2018twin,ma2018phase,yin2019measurement,lin2018simple,cui2019twin,curty2019simple,yin2019coherent,maeda2019repeaterless} can extract the secret key that surpasses the Pirandola-Laurenza-Ottaviani-Banchi (PLOB) bound~\cite{pirandola2017fundamental} by introducing an intermediate station to perform single-photon-type interference. Recently, using twin field theory~\cite{lucamarini2018overcoming}, one differential-phase-shift QSS has been proposed to provide security against individual attacks~\cite{Gu:21}, followed by analysis of finite-key effects and the asymmetric regime~\cite{e23060716}. Here, we present an unconditionally secure three-user QSS protocol without signal disturbance monitoring for outside and inside attackers. Our new QSS scheme is the first realistic extended application with the encoding technique that spreads quantum information coherently over hundreds of laser pulses~\cite{sasaki2014practical}. The natures of high noise tolerance and irrelevance to signal disturbance still remain in this QSS protocol, which allows to generate secure key even the bit error rate is up close to $50\%$ for outside adversaries. Our protocol only requires weak coherent sources and single photon detectors. Numerical simulations show that it can be theoretically implemented over 600 km using recently developed techniques in practical quantum key distribution~\cite{takesue2015experimental,wang2015experimental,yin2018improved,minder2019experimental,zhong2019proof,wang2019beating,chen2020sending,fang2020implementation}.

\section{PROTOCOL DESCRIPTION}
\begin{figure}[htbp]
\centering
\includegraphics[width=8.6cm]{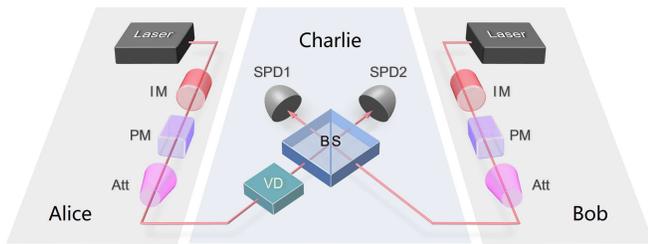}
\caption{Practical implementation of QSS. Alice and Bob utilize the continuous-wave laser to prepare the phase stabilized continuous light. They employ intensity modulator (IM) to generate the periodic laser pulse. For each pulse train, Alice and Bob exploit the phase modulator (PM) to apply phase shift $\{0,\pi\}$ on each pulse according to the random bit strings $\{s_{j_{A}}\}$ and $\{s_{i_{B}}\}$, respectively. The attenuator (Att) is used to implement weak pulse with single photon level modulation. Charlie exploits a variable delay (VD) to change the arrived time of Alice's pulse train with earlier or later $(-1)^{b}r$ period. He then superposes the two pulse trains using a beam splitter (BS) and compares the phase difference.
} \label{setup}
\end{figure}

The schematic of QSS that how to overcome the PLOB bound is shown in Fig.~\ref{setup}. Our QSS protocol has a readiness for building a star network, where two symmetric remote players, Alice and Bob are connected to the central dealer, Charlie. Charlie implements an interference measurement on the two laser pulses sent by Alice and Bob, which will reveal that the corresponding phases are equal or different between Alice and Bob and thus forms the bit value correlation as for QSS. The raw key rate scales with the square-root $O(\sqrt{\eta})$ of the total channel transmittance (between Alice and Bob) due to the single-photon-type interference of Charlie. The photon detected by Charlie can be considered from twin fields coming from both Alice and Bob, respectively. In order to resist attacks of a dishonest player or the outside eavesdropper, Charlie introduces two random numbers $r\in\{1,\ldots,L-1\}$ and $b\in\{0,1\}$ to post-selected the effective bit as raw key. No one can access the random numbers $r,~b$ in advance, which indicates that the leaked information is ignorable.

To be precise, our QSS protocol is inspired by the recently developed round-robin differential phase-shift quantum key distribution~\cite{sasaki2014practical} and twin-field quantum key distribution~\cite{lucamarini2018overcoming}. Here, we name this protocol as round-robin (RR) QSS, which proceeds as follows:

(i) For each train, Alice and Bob independently and randomly generate bit strings $\{s_{j_{A}}\}$ and $\{s_{i_{B}}\}$ with $j_{A},i_{B}\in\{1,2,\ldots,L\}$ and $s_{j_{A}},s_{i_{B}}\in\{0,1\}$. They both send a train of $L$ laser pulses to Charlie simultaneously, with each pulse randomly added a phase shift $0$ or $\pi$ according to their random bits. The corresponding quantum states of Alice and Bob can be represented as
\begin{equation}\label{Astate}
\begin{aligned}
\ket{\Psi_{A}}&=\bigotimes_{j_{A}=1}^{L}\ket{e^{is_{j_{A}}\pi}\sqrt{\mu/L}}_{j_{A}},
\end{aligned}
\end{equation}
and
\begin{equation}\label{Bstate}
\begin{aligned}
\ket{\Psi_{B}}&=\bigotimes_{i_{B}=1}^{L}\ket{e^{is_{i_{B}}\pi}\sqrt{\mu/L}}_{i_{B}},
\end{aligned}
\end{equation}
where $\ket{e^{is_{j_{A}}\pi}\sqrt{\mu/L}}_{j_{A}}$ designates the $j_{A}$th laser pulse of Alice and $\mu$ is the total intensity of $L$ optical pulses.

(ii) Charlie will let the received two laser pulse trains interfere after he implements a random operation on Alice's laser pulses. The random operation is that Charlie changes the arrival time of Alice's pulse train related to that of Bob's with
\begin{equation}\label{eq3}
\begin{aligned}
j_{A}-i_{B}=(-1)^{b}r,
\end{aligned}
\end{equation}
which means that the $j_{A}=i_{B}+(-1)^{b}r$th laser pulse of Alice and $i_{B}$th laser pulse of Bob will interfere.

(iii) Let one and only one photon click from interference measurements and no other click in the whole pulse train denote an effective detection event. Charlie records his raw key bit $X_{C}$ as 0 and 1 for the phase difference with 0 and $\pi$ given by an effective detected event.

(iv) Charlie announces $j_{A}$ and $i_{B}$ to Alice and Bob through the authenticated classical channel when the effective detection event is acquired from the superposed $j_{A}$ and $i_{B}$ pulses.

(v) Alice and Bob record $X_{A}=s_{j_{A}}$ and $X_{B}=s_{i_{B}}$ as their raw key bits.

(vi) All users (Alice, Bob and Charlie) exploit the error correction, error verification and privacy amplification of multiparty scheme to obtain the secure key.

\begin{figure}[htbp]
\centering
\includegraphics[width=8.6cm]{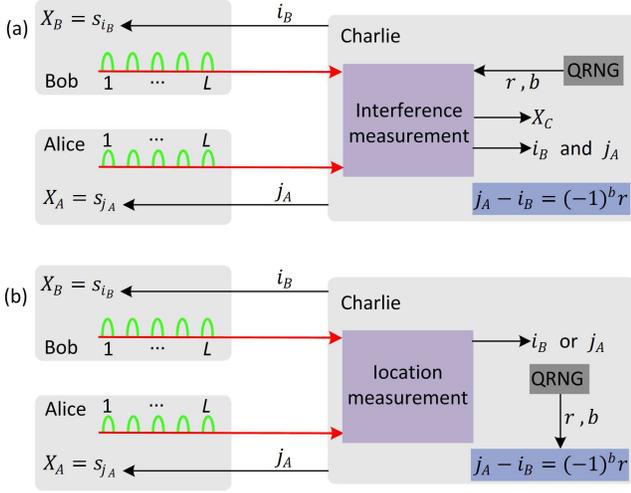}
\caption{Charlie's alternative choices of measurements. Alice and Bob send the quantum signals to Charlie via insecure quantum channel. The raw key bit held by Alice (Bob) is $X_{A}=s_{j_{A}}$ ($X_{B}=s_{i_{B}}$), where $j_{A}$ ($i_{B}$) is published by Charlie. The dishonest player, Alice or Bob, and the outside Eve try to obtain the raw key $X_{A}\oplus X_{B}$ in both figures. (a) Charlie exploits the interference measurement with random numbers $\{r,b\}$ to acquire raw key bit $X_{C}$ that is equal to $X_{A}\oplus X_{B}$ in the ideal case. (b)
Charlie uses the random numbers $\{r,b\}$ to generate $j_{A}$ ($i_{B}$) with the relationship $j_{A}-i_{B}=(-1)^{b}r$ after he utilizes the location measurement to obtain $i_{B}$ ($j_{A}$). All attacks by the adversary should have the same results in both figures since the announced values $\{j_{A},i_{B}\}$ are identical in figures (a) and (b).
} \label{virtual}
\end{figure}

\section{SECURITY ANALYSIS}
In order to provide an intuitive understanding of the proposed RRQSS protocol, we demonstrate its security by
considering two equivalent cases shown in Fig.~\ref{virtual}, which is the generalization of the security proof method of round-robin differential phase-shift quantum key distribution~\cite{sasaki2014practical}. Here, we assume that the interference measurement and location measurement can discriminate the detection signal exactly coming from a single photon. Another assumption we have to make is that the devices used by players and the dealer are perfect and leak no more information to eavesdroppers. These assumptions are similar to those in the round-robin differential phase-shift quantum key distribution~\cite{sasaki2014practical} and the first assumption can be solved by using the detector-decoy method~\cite{yin2016detector}.
Charlie announces a successful detection when one and only one photon is clicked in both Fig.~\ref{virtual}a and Fig.~\ref{virtual}b.

The setup of Fig.~\ref{setup} is a practical implementation of scheme in Fig.~\ref{virtual}a with efficiency 1/2. In the ideal case, the raw key of Charlie in Fig.~\ref{virtual}a is $X_{C}=X_{A}\oplus X_{B}$, where $X_{A}=s_{j_{A}}$ and $X_{B}=s_{i_{B}}$ are the bit values of Alice and Bob, respectively. It means that Alice and Bob should collaborate to access the information of Charlie.
It has been proven that the classical bits of information that can be transmitted by sending the coherent fingerprint state of Eqs.~\eqref{Astate} and \eqref{Bstate} satisfies $O(\mu \log_{2}L)$~\cite{Arrazola2014Quantum,Guan2016Observation}. It is much less than the encoding bit information of Alice or Bob for each pulse train with large $L$.
At least one of $i_{B}$ and $j_{A}$ cannot be confirmed due to $j_{A}-i_{B}=(-1)^{b}r$ with random numbers $\{r,b\}$ before Charlie implements measurement, which means anyone can have only a little knowledge of the raw key $X_{C}$ of Charlie or bit values $X_{A}\oplus X_{B}$, including the dishonest player or the outside eavesdropper Eve.
In order to access how much the adversary knows about the bit values $X_{A}\oplus X_{B}$, we imagine that Charlie exploits the alternative choice with the location measurement in Fig.~\ref{virtual}b. Charlie directly acquires $i_{B}$ or $j_{A}$ using the location measurement ($i_{B}$ or $j_{A}$). Then, he exploits random numbers $\{r,b\}$ to generate the other index satisfying $j_{A}-i_{B}=(-1)^{b}r$. To be more specific, Charlie's interference measurement is characterized by a set of projection measurement operators
\begin{equation}
\begin{aligned}
\hat{E}_{k,s}^{r,b}=\hat{P}\left(\frac{\ket{k}_{B}+(-1)^{s}\ket{k+(-1)^{b}r}_{A}}{\sqrt{2}}\right),
\end{aligned}
\end{equation}
where $\hat{P}(\ket{\varphi})=\ket{\varphi}\bra{\varphi}$, $k\in\{1,\ldots,d\}$, $k+(-1)^{b}r\in\{1,\ldots,d\}$ and $s\in\{0,1\}$. State $\ket{k}$ denotes the photon in the $k$th pulse.
For single-photon input state $\hat{\rho}$, the probability of output $\{k,s\}$ is given by ${\rm Tr}(\hat{\rho}\hat{E}_{k,s}^{r,b})/2$, since there exists a filter with efficiency $1/2$~\cite{sasaki2014practical}. We remark that the state $\hat{\rho}$ is the joint quantum state between Alice and Bob. Charlie publishes indices $\{k_{B}, [k_{B}+(-1)^{b}r]_{A}\}$ and acquires $X_{C}=s$ from this output. Therefore, the probability of announcing $\{j_{A},i_{B}\}$ ($j_{A}=i_{B}+(-1)^{b}r$) is written as
\begin{equation}
\begin{aligned}
p(\{j_{A},i_{B}\})=[p(j_{A})+p(i_{B})]/2,
\end{aligned}
\end{equation}
where $p(k)=\langle k|\hat{\rho}|k\rangle$ is the probability of detecting a photon in the $k$th pulse. And Charlie's location measurement is characterized by a set of projection measurement operators
\begin{equation}
\begin{aligned}
\hat{E'}_{k}=\frac{1}{2}[\hat{P}(\ket{k}_{B})+\hat{P}(\ket{k}_{A})]
\end{aligned}
\end{equation}
where $k\in\{1,\ldots,d\}$. Thereby, the probability of announcing $\{j_{A},i_{B}\}$ ($j_{A}=i_{B}+(-1)^{b}r$) can be given by
\begin{equation}
\begin{aligned}
p'(\{j_{A},i_{B}\})=[p(j_{A})+p(i_{B})]/2.
\end{aligned}
\end{equation}
Obviously, $p(\{j_{A},i_{B}\})=p'(\{j_{A},i_{B}\})$, it means that the published indices $\{j_{A},i_{B}\}$ are identical for two figures in Fig.~\ref{virtual}. And the knowledge of bit value $X_{A}\oplus X_{B}$ in actual and virtual protocols are equivalent for anyone except for Charlie, which indicates the generation procedure of $\{j_{A},i_{B}\}$ is equivalent to the interference measurement in Fig.~\ref{virtual}a. Therefore, the adversary have the same knowledge of $X_{A}\oplus X_{B}$ in Fig.~\ref{virtual}a and Fig.~\ref{virtual}b since she or he cannot distinguish the choices of measurement by Charlie.

In the RRQSS protocol, at most one player is dishonest. We first demonstrate that the protocol is secure for outside Eve and provide the detailed secret key rate. Then we will verify the security against one dishonest player, Alice or Bob and the corresponding key rate.

\subsection{The outside Eve}
First, we consider the case where Alice, Bob and Charlie are all honest and collaborate to generate the secret key with $X_{C}=X_{A}\oplus X_{B}$. The outside Eve is an eavesdropper, who tries to attack the knowledge of $X_{A}\oplus X_{B}$. Obviously, this RRQSS protocol is similar with the round-robin differential phase-shift quantum key distribution~\cite{sasaki2014practical} if we consider Alice and Bob as a single user.
In order to calculate the secret key rate, we employ the source-replacement scheme~\cite{sasaki2014practical}. We assume that Alice prepares an entangled state between $L$ virtual qubit and $L$ optical pulses instead of a train of coherent state,
\begin{equation}
\begin{aligned}\label{eq4}
2^{-L/2}\bigotimes_{j_{A}=1}^{L}\sum_{s_{j_{A}}=0,1}\ket{s_{j_{A}}}_{A}\ket{e^{is_{j_{A}}\pi}\sqrt{\mu/L}}_{j_{A}},
\end{aligned}
\end{equation}
where $\ket{s_{j_{A}}}_{A}$ is the eigenstate of the $Z$ basis in the $j_{A}$th virtual qubit of Alice. Similarly, Bob generates an entangled state between $L$ virtual qubit and $L$ optical pulses,
\begin{equation}
\begin{aligned}\label{eq5}
2^{-L/2}\bigotimes_{i_{B}=1}^{L}\sum_{s_{i_{B}}=0,1}\ket{s_{i_{B}}}_{B}\ket{e^{is_{i_{B}}\pi}\sqrt{\mu/L}}_{i_{B}},
\end{aligned}
\end{equation}
where $\ket{s_{i_{B}}}_{B}$ is the eigenstate of the $Z$ basis in the $i_{B}$th virtual qubit of Bob. Note that the source-replacement scheme does not change security. Since the $L$ optical pulses sent by Alice or Bob are identical to those in the actual scheme if Alice and Bob measure the virtual qubit in the $Z$ basis before sending optical pulses.

The secret key rate per pulse of the RRQSS in the case of the outside Eve is given by
\begin{equation}
\begin{aligned}\label{eq6}
R=\frac{Q}{L}[1-h(e_{\rm p})-fh(e_{\rm b})],
\end{aligned}
\end{equation}
where $Q$ is the gain of transmitting $L$ pulse pair trains, $e_{\rm b}$ and $e_{\rm p}$ are the bit and phase error rates of this protocol related to error correction and privacy amplification, respectively. A bit error is defined as $X_{A}\oplus X_{B}\neq X_{C}$. Specifically, the average gain $Q$ and bit error rate $e_{\rm b}$ of each train can be written as
\begin{equation}
\begin{aligned}
Q=\frac{1}{2}[1-(1-Lp_{d})e^{-2\mu \sqrt{\eta}}],
\end{aligned}
\end{equation}
and
\begin{equation}
\begin{aligned}
e_{\rm b}=\frac{e_{d}(1-e^{-2\mu \sqrt{\eta}})+Lp_{d}e^{-2\mu \sqrt{\eta}}/2}{1-(1-Lp_{d})e^{-2\mu \sqrt{\eta}}},
\end{aligned}
\end{equation}
where $\sqrt{\eta}=\eta_{d}\times10^{-\alpha D/20}$ is the efficiency between Alice (Bob) and Charlie. $D$ is the total distance among Alice, Bob and Charlie.

Let $\nu$ be the total photon number in a train of optical pulses with total intensity $\mu$. Then, the probability of finding more than $\nu_{\rm th}$ photons in a train of optical pulses can be written as
\begin{equation}
\begin{aligned}
{\rm Pr}(\nu>\nu_{\rm th})=e_{\rm src}:=1-\sum_{\nu=0}^{\nu_{\rm th}}\frac{e^{-\mu}\mu^{\nu}}{\nu!},
\end{aligned}
\end{equation}
where $\nu_{\rm th}$ is an integer constant chosen in this protocol.

Considering a fictitious situation in Fig.~\ref{virtual}b, Alice and Bob respectively deliver these virtual qubits to Charlie with a secure manner. Charlie simply applies a controlled-NOT operation to the $i_{B}$th and $j_{A}$th virtual qubits, with the $i_{B}$th ($j_{A}$th) virtual qubit as the control and the $j_{A}$th ($i_{B}$th) virtual qubit as the target given $i_{B}$ ($j_{A}$), $b$ and $r$. The bit value $X_{A}\oplus X_{B}=s_{i_{B}}\oplus s_{j_{A}}$ becomes the outcome of $Z$-basis measurement on the $j_{A}$th ($i_{B}$th) target virtual qubit. Let $\ket{\pm}$ be the eigenstates of $X$-basis. Note that the controlled-NOT operation will not affect the $X$-basis eigenstate of the target virtual qubit~\cite{sasaki2014practical}. If the target virtual qubit is fixed to state $\ket{+}$, there is no knowledge of the measurement outcome on the $Z$-basis acquired by Eve. The phase error rate is defined as the probability of finding the target virtual qubit in state $\ket{-}$~\cite{Koashi2009simple}. From the Eqs. \eqref{eq4} and \eqref{eq5}, one can find out that the virtual qubit will be state $\ket{+}$ ($\ket{-}$) for even (odd) number photons in each optical pulse.  Therefore, the probability of a phase error is bounded by $\nu_{\rm th}/(L-1)$ when the number of photons in this train is no more than $\nu_{\rm th}$. Note that the target virtual qubit is randomly selected from $L-1$ virtual qubits.
The phase error rate of the RR-QSS protocol for the outside Eve can be given by
\begin{equation}
\begin{aligned}\label{eq8}
e_{\rm p}=\frac{2e_{\rm src}}{Q}+\left(1-\frac{2e_{\rm src}}{Q}\right)\frac{\nu_{\rm th}}{L-1}.
\end{aligned}
\end{equation}
The first term of Eq.~\eqref{eq8} corresponds to the fraction of a sifted key that the number of contained photons in a train pulses (both for honest Alice and honest Bob) are larger than $\nu_{\rm th}$ and should be regarded as a phase error in the worst-case scenario.

\begin{table}
\centering
\caption{Simulation parameters~\cite{chen2020sending}. $\eta_{d}$ and $p_{d}$ are the detector efficiency and dark count rate. $e_{d}$ is the misalignment rate. $\alpha$ is the attenuation coefficient of the ultralow-loss fiber. $f$ is the error correction inefficiency.}\label{tab1}
\begin{tabular}{ccccc}
\hline
\hline
$\eta_{d}$ ~~~~~& $p_{d}$ ~~~~~& $e_{d}$~~~~~ & $\alpha$ ~~~~~& $f$\\
\hline
$56\%$~~~~~& $10^{-8}$~~~~~ & $2\%$ ~~~~~ & $0.167$ ~~~~~& $1.1$\\
\hline
\hline
\end{tabular}
\end{table}

\begin{figure}[htbp]
\centering
\includegraphics[width=8.6cm]{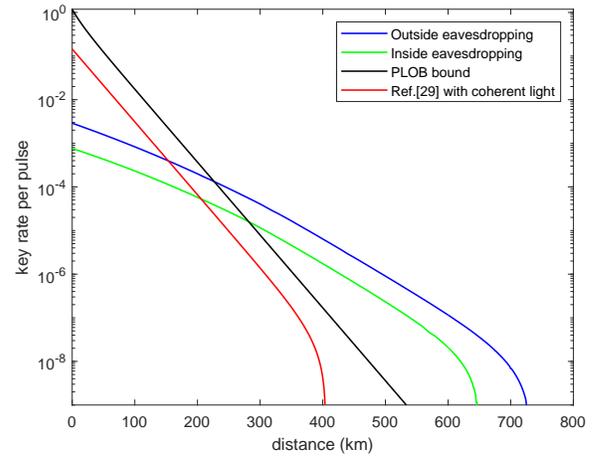}
\caption{Final key rate (per pulse) in logarithmic scales as a function of the total distance among three users with $e_{d} = 2\%$. The performance of our protocol and single-qubit QSS is measured in terms of the key rate per pulse $R$. With parameters in Table.~\ref{tab1}, the key rates of our protocol against outside and inside eavesdropping are optimized with $\mu$, $\nu_{th}$ and $L$ and the key rate in~\cite{schmid2005experimental} is optimized with $\mu$, together with the assumption of weak coherent light source and asymptotic condition of infinite decoy states.
} \label{inout}
\end{figure}

\subsection{The inside Alice or Bob}
Then, we consider the case where Alice, Bob and Charlie collaborate to generate the secret key with $X_{C}=X_{A}\oplus X_{B}$ with one player, Alice or Bob dishonest. The inside eavesdropper tries to attack the knowledge of $X_{A}\oplus X_{B}$ with the help of the outside Eve.

Here, we first consider Bob to be the adversary. The bit value $X_{B}$ is all known by Bob, which means that he only needs to obtain the information of $X_{A}$. According to the above security analysis, there are two cases in Fig.~\ref{virtual}b where the location measurement generates $i_{B}$ or $j_{A}$.
For the $i_{B}$ click case, the bit value $X_{A}\oplus X_{B}$ is acquired by directly measuring the target qubit of honest Alice with $Z$ basis. The corresponding phase error rate is $e_{{\rm p}_{i_{B}}}=\frac{e_{\rm src}}{Q_{B}}+\left(1-\frac{e_{\rm src}}{Q_{B}}\right)\frac{\nu_{\rm th}}{L-1}$. Here, $Q_{B}$ is probability of $i_{B}$ generated in location measurement and $e_{\rm src}$ is calculated using the honest Alice's optical pulses. For the $j_{A}$ click case, one cannot directly measure the phase error on target qubit of dishonest Bob since the intensity and photon number distribution sent by Bob are uncertain. Considering the worst case, Bob will acquire all information about the $j_{A}$ click case.

Now we consider Alice to be the adversary. For the $j_{A}$ click case, the phase error rate is $e_{{\rm p}_{j_{A}}}=\frac{e_{\rm src}}{Q_{A}}+\left(1-\frac{e_{\rm src}}{Q_{A}}\right)\frac{\nu_{\rm th}}{L-1}$. Here, $Q_{A}$ is probability of $j_{A}$ generated in location measurement and $e_{\rm src}$ is calculated by using the honest Bob's optical pulses characteristic. Obviously, we have $Q=Q_{A}+Q_{B}$.  For the $i_{B}$ click case, Alice will obtain all information.

For cases of the inside adversary, no one can know in advance whether the adversary is  Alice or Bob. Therefore, the secret key rate per pulse of the RRQSS in the case of the inside adversary can be written as
\begin{equation}\label{eq9}
\begin{aligned}
R=\frac{1}{L}\{\hat{Q}[1-h(\hat{e}_{\rm p})]-Qfh(e_{\rm b})\},
\end{aligned}
\end{equation}
where we have $\hat{Q}=\min\{Q_{A},Q_{B}\}$ and phase error rate
\begin{equation}\label{eq10}
\begin{aligned}
\hat{e}_{p}=\frac{e_{\rm src}}{\hat{Q}}+\left(1-\frac{e_{\rm src}}{\hat{Q}}\right)\frac{\nu_{\rm th}}{L-1}.
\end{aligned}
\end{equation}

\section{PERFORMANCE}
Here, we display the secret key rate per pulse $R$ of our protocol, assuming that Charlie holds a quantum random number generator to generate $r$ and $b$ in Eq.~\eqref{eq3}. Without loss of generality, we have the gains $Q_{A}=Q_{B}=Q/2$ due to the symmetric positions of Alice and Bob. Our protocol only needs two continuous-wave laser to prepare phase-stabilized continuous light with total intensity $\mu/L$ and phase $0$ or $\pi$, which is available to current techniques. At Charlie's site, single-photon detectors are assumed with the same efficiency $\eta_d$ and the dark count $p_d$. We introduce $\alpha$ as the attenuation coefficient of the ultra low-loss fiber and $f$ as the error correction inefficiency. For simplicity, we assume a misalignment error rate $e_d$ to generalize the whole systematic error rate in our protocol. Specific value of parameters is listed in Table.~\ref{tab1}. Additionally, to illustrate that the key rate of our protocol scales with the square-root of the total channel transmittance, the PLOB bound between Alice and Bob is also considered for comparison ($R_{\rm PLOB}= -\log_{2}(1-\eta)$ and $\eta=\eta_d \times 10^{-\alpha D/10}$). We also consider the key rate of single-qubit QSS~\cite{schmid2005experimental}, which is another practical QSS protocol with several experimental implementations. For simulation, we assume that the weak coherent light source and the asymptotic condition of infinite decoy states are applied in single-qubit QSS.

\begin{figure}[htbp]
\centering
\includegraphics[width=8.6cm]{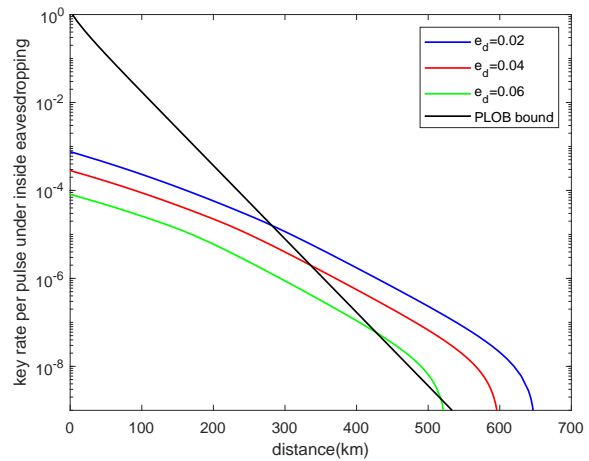}
\caption{Key rate (per pulse) in logarithmic scales as a function of the total distance among three-user with different misalignment rates.
The key rate of our QSS protocol can also surpass the PLOB bound even when the misalignment rate is more than $e_{d}=6\%$.
} \label{rrtfqss}
\end{figure}

Moreover, we can take a little further step to consider the expected behavior of finite-sized key rate in our protocol using the same idea in round-robin differential phase-shift quantum key distribution~\cite{sasaki2014practical}. Let $N$ be the number of rounds and first we define the tail distribution for finding more than $a$ successful events in a binomial distribution as $\bar{f}(a;~n,p) = \sum_{k>a}p^k(1-p)^{n-k}n!/[k!(n-k)!]$. Then the probability of choosing $Nr_1$ from $N$ sifted bits to include all the tagged portion will be $1-\epsilon_1$, where $\epsilon_1 = \bar{f}(Nr_1;~N_{\rm round},e_{src})$ and $N_{\rm round}$ denotes the number of rounds of transmitting $L$ pulses. Here we assume there is no phase error in the chosen $Nr_1$ bits. When considering phase errors in the remaining $N'$ bits ($N' = N(1-r_1)$), it should be less than $N'r_2$ for a probability $1-\epsilon_2$ and $\epsilon_2 = \bar{f}(N'r_2;~N',\nu_{th}/(L-1))$. With $\epsilon_1$ and $\epsilon_2$, we can characterize the imperfection in the final key given the bits of information leakage to the eavesdropper. Given $s>0$, we let $\epsilon_1=\epsilon_2=2^{-s}$ by choosing proper $r_1$ and $r_2$. Then, we have $h(\hat{e}_p)=r_1 + (1-r_1)h(r_2)+s/N$ with $s$ commonly ranging from 70 to 160. With a crude Gaussian approximation of ${\rm ln}[\bar{f}(a;n,p)] \cong -(a-np)^2/[2np(1-p)]$, we have that $r_1 \cong p_1 + \sqrt{(2 {\rm ln2})p_1(s/N)}$ and $r_2 \cong p_2 + \sqrt{(2 {\rm ln2})p_2(1-p_2)(s/N)}$ where $p_1 = e_src/\hat{Q}$ and $p_2 = \nu_{th}/(L-1)$. For $N \gg s$, substituting numerics gives $h(\hat{e}_p) \cong h^{asy}(\hat{e}_p)(1+1.98\sqrt{s/N})$ and $h^{asy}(\hat{e}_p)= p_1 + (1-p_1)h(p_2)$.

\begin{figure}[htbp]
\centering
\includegraphics[width=8.6cm]{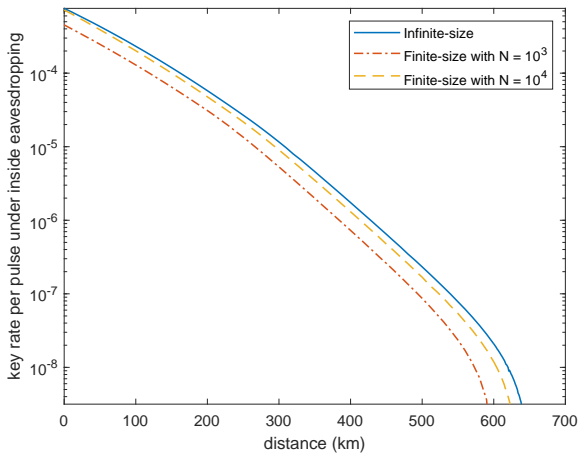}
\caption{Infinite-size key rate (per pulse) versus finite-size key rate (per pulse) under inside eavesdropping. We fix that $s=100$ and plot the finite-size key rate with $N=10^3$ (dash-dot line) and $N=10^4$ (dashed line).
} \label{finite}
\end{figure}

With the above assumptions, the performance of our QSS protocol is presented in Figs.~\ref{inout},~\ref{rrtfqss} and~\ref{finite} by numerically optimizing the secret key rate over the free parameters $\mu$, $\nu_{\rm th}$ and $L$. As depicted in Fig.~\ref{inout}, besides the key rate of our QSS protocol under inside and outside eavesdropping, we also plot the PLOB bound and the key rate in single-qubit QSS~\cite{schmid2005experimental} where the misalignment rate $e_d=2\%$. Evidently, our QSS protocol has better performance over single-qubit QSS and can break the PLOB bound for $D$ larger than $\sim$ 300 km with the final transmission distance approaching over 600 km. In addition, considering impact on the final key rate caused by the misalignment error rate $e_d$, we present the key rate of our QSS protocol with different misalignment rate in Fig.\ref{rrtfqss}. Although the misalignment rate $e_d$ is more than $6\%$, the final key rate can also surpass the linear PLOB bound, which shows the high tolerance of errors in our protocol. With $e_d$ smaller than $4\%$, the theoretical transmission distance will approach over 600 km, which is suitable to intercity multiparty quantum communication. Fig.~\ref{finite} shows that there is little influence the finite-size analysis has on the final key rate of our protocol. When $N=10^4$, the transmission distance will also approach over 600 km.

\section{CONCLUSION}
In summary, we present a practical QSS protocol containing three users with unconditional security. Inspired by the twin-field quantum key distribution, the key rate scales as $O(\sqrt{\eta})$ rather than $O(\eta)$, breaking the PLOB bound introduced by capacity of quantum channels~\cite{takeoka2014fundamental,pirandola2017fundamental}. Moreover, we propose the security analysis to demonstrate the unconditional security of our protocol under both outside and inside eavesdropping. The key point here is to consider eavesdropping of the inside participant Alice or Bob, which is quite different from the original round-robin quantum key distribution. Namely, the security is irrelevant to the channel we use. In addition, although constricted by the inside eavesdropper, our protocol still shows a long theoretical transmission distance of over 600 km under small misalignment error rate, which reveals the potential for long distance multiparty communication with high efficiency. According to the recent work~\cite{bouchard2018round}, RRQSS can also be applied in the experimental and practical implementations using the orbital angular momentum degree of freedom. Note that our QSS protocol shares classical bits rather than quantum states \cite{Cleve:1999:How}.

Furthermore, the apparatus settings in our scheme can be directly deployed to realize the third-man quantum cryptography (TQC)~\cite{zukowski1998quest} where Charlie plays the role of a controller to determine whether to announce his measurement results. If Charlie announces his measurement results, Alice's bits and Bob's bits will build up their own perfect correlations. Then based on the correlations, Alice and Bob will create the secret key and Charlie has no knowledge of the keys which Alice and Bob have created. Namely, in the TQC, Alice and Bob will fail to generate the secret key without the help of the controller, Charlie~\cite{chen2005experimental}. The secret key rate of our TQC protocol can be given by the Eq.\eqref{eq6}. In our TQC scheme, we have to consider Charlie is an honest participant but he has no knowledge of secret key. Therefore, the practical security of the TQC here is between twin-field~\cite{lucamarini2018overcoming} and trusted relay~\cite{chen2021integrated} quantum key distribution.

Our RRQSS protocol is a first step to extend the application of round-robin method from quantum key distribution~\cite{sasaki2014practical}. Its security analysis guarantees that our QSS protocol is also independent of any intervention by eavesdroppers and has high tolerance of noise. Future work is necessary to narrow the gap of secret key rate between inside and outside eavesdropping in Fig.~\ref{inout}. Combined with long transmission distance and simple apparatus requirements, we believe our protocol is suitable for secure multiparty communication in the future quantum networks.

\section*{FUNDING}
National Natural Science Foundation of China (61801420); Key-Area Research and Development Program of Guangdong Province (2020B0303040001); Fundamental Research Funds for the Central Universities (020414380182).

%\bibliographystyle{apsrev4-1}
%%%%%%%%%%% If using BibTeX:
%\bibliography{Bibli}

\begin{thebibliography}{66}%
\makeatletter
\providecommand \@ifxundefined [1]{%
 \@ifx{#1\undefined}
}%
\providecommand \@ifnum [1]{%
 \ifnum #1\expandafter \@firstoftwo
 \else \expandafter \@secondoftwo
 \fi
}%
\providecommand \@ifx [1]{%
 \ifx #1\expandafter \@firstoftwo
 \else \expandafter \@secondoftwo
 \fi
}%
\providecommand \natexlab [1]{#1}%
\providecommand \enquote  [1]{``#1''}%
\providecommand \bibnamefont  [1]{#1}%
\providecommand \bibfnamefont [1]{#1}%
\providecommand \citenamefont [1]{#1}%
\providecommand \href@noop [0]{\@secondoftwo}%
\providecommand \href [0]{\begingroup \@sanitize@url \@href}%
\providecommand \@href[1]{\@@startlink{#1}\@@href}%
\providecommand \@@href[1]{\endgroup#1\@@endlink}%
\providecommand \@sanitize@url [0]{\catcode `\\12\catcode `\$12\catcode
  `\&12\catcode `\#12\catcode `\^12\catcode `\_12\catcode `\%12\relax}%
\providecommand \@@startlink[1]{}%
\providecommand \@@endlink[0]{}%
\providecommand \url  [0]{\begingroup\@sanitize@url \@url }%
\providecommand \@url [1]{\endgroup\@href {#1}{\urlprefix }}%
\providecommand \urlprefix  [0]{URL }%
\providecommand \Eprint [0]{\href }%
\providecommand \doibase [0]{http://dx.doi.org/}%
\providecommand \selectlanguage [0]{\@gobble}%
\providecommand \bibinfo  [0]{\@secondoftwo}%
\providecommand \bibfield  [0]{\@secondoftwo}%
\providecommand \translation [1]{[#1]}%
\providecommand \BibitemOpen [0]{}%
\providecommand \bibitemStop [0]{}%
\providecommand \bibitemNoStop [0]{.\EOS\space}%
\providecommand \EOS [0]{\spacefactor3000\relax}%
\providecommand \BibitemShut  [1]{\csname bibitem#1\endcsname}%
\let\auto@bib@innerbib\@empty
%</preamble>
\bibitem [{\citenamefont {Kimble}(2008)}]{kimble2008quantum}%
  \BibitemOpen
  \bibfield  {author} {\bibinfo {author} {\bibfnamefont {H.~J.}\ \bibnamefont
  {Kimble}},\ }\href@noop {} {\bibfield  {journal} {\bibinfo  {journal}
  {Nature}\ }\textbf {\bibinfo {volume} {453}},\ \bibinfo {pages} {1023}
  (\bibinfo {year} {2008})}\BibitemShut {NoStop}%
\bibitem [{\citenamefont {Wehner}\ \emph {et~al.}(2018)\citenamefont {Wehner},
  \citenamefont {Elkouss},\ and\ \citenamefont {Hanson}}]{wehner2018quantum}%
  \BibitemOpen
  \bibfield  {author} {\bibinfo {author} {\bibfnamefont {S.}~\bibnamefont
  {Wehner}}, \bibinfo {author} {\bibfnamefont {D.}~\bibnamefont {Elkouss}}, \
  and\ \bibinfo {author} {\bibfnamefont {R.}~\bibnamefont {Hanson}},\
  }\href@noop {} {\bibfield  {journal} {\bibinfo  {journal} {Science}\ }\textbf
  {\bibinfo {volume} {362}},\ \bibinfo {pages} {eaam9288} (\bibinfo {year}
  {2018})}\BibitemShut {NoStop}%
\bibitem [{\citenamefont {Ladd}\ \emph {et~al.}(2010)\citenamefont {Ladd},
  \citenamefont {Jelezko}, \citenamefont {Laflamme}, \citenamefont {Nakamura},
  \citenamefont {Monroe},\ and\ \citenamefont {O'Brien}}]{ladd2010quantum}%
  \BibitemOpen
  \bibfield  {author} {\bibinfo {author} {\bibfnamefont {T.~D.}\ \bibnamefont
  {Ladd}}, \bibinfo {author} {\bibfnamefont {F.}~\bibnamefont {Jelezko}},
  \bibinfo {author} {\bibfnamefont {R.}~\bibnamefont {Laflamme}}, \bibinfo
  {author} {\bibfnamefont {Y.}~\bibnamefont {Nakamura}}, \bibinfo {author}
  {\bibfnamefont {C.}~\bibnamefont {Monroe}}, \ and\ \bibinfo {author}
  {\bibfnamefont {J.~L.}\ \bibnamefont {O'Brien}},\ }\href@noop {} {\bibfield
  {journal} {\bibinfo  {journal} {Nature}\ }\textbf {\bibinfo {volume} {464}},\
  \bibinfo {pages} {45} (\bibinfo {year} {2010})}\BibitemShut {NoStop}%
\bibitem [{\citenamefont {Arute}\ \emph {et~al.}(2019)\citenamefont {Arute},
  \citenamefont {Arya}, \citenamefont {Babbush}, \citenamefont {Bacon},
  \citenamefont {Bardin}, \citenamefont {Barends}, \citenamefont {Biswas},
  \citenamefont {Boixo}, \citenamefont {Brandao}, \citenamefont {Buell},
  \citenamefont {Burkett}, \citenamefont {Chen}, \citenamefont {Chen},
  \citenamefont {Chiaro}, \citenamefont {Collins}, \citenamefont {Courtney},
  \citenamefont {Dunsworth}, \citenamefont {Farhi}, \citenamefont {Foxen},
  \citenamefont {Fowler}, \citenamefont {Gidney}, \citenamefont {Giustina},
  \citenamefont {Graff}, \citenamefont {Guerin}, \citenamefont {Habegger},
  \citenamefont {Harrigan}, \citenamefont {Hartmann}, \citenamefont {Ho},
  \citenamefont {Hoffmann}, \citenamefont {Huang}, \citenamefont {Humble},
  \citenamefont {Isakov}, \citenamefont {Jeffrey}, \citenamefont {Jiang},
  \citenamefont {Kafri}, \citenamefont {Kechedzhi}, \citenamefont {Kelly},
  \citenamefont {Klimov}, \citenamefont {Knysh}, \citenamefont {Korotkov},
  \citenamefont {Kostritsa}, \citenamefont {Landhuis}, \citenamefont
  {Lindmark}, \citenamefont {Lucero}, \citenamefont {Lyakh}, \citenamefont
  {Mandr¨¤}, \citenamefont {McClean}, \citenamefont {McEwen}, \citenamefont
  {Megrant}, \citenamefont {Mi}, \citenamefont {Michielsen}, \citenamefont
  {Mohseni}, \citenamefont {Mutus}, \citenamefont {Naaman}, \citenamefont
  {Neeley}, \citenamefont {Neill}, \citenamefont {Niu}, \citenamefont {Ostby},
  \citenamefont {Petukhov}, \citenamefont {Platt}, \citenamefont {Quintana},
  \citenamefont {Rieffel}, \citenamefont {Roushan}, \citenamefont {Rubin},
  \citenamefont {Sank}, \citenamefont {Satzinger}, \citenamefont {Smelyanskiy},
  \citenamefont {Sung}, \citenamefont {Trevithick}, \citenamefont
  {Vainsencher}, \citenamefont {Villalonga}, \citenamefont {White},
  \citenamefont {Yao}, \citenamefont {Yeh}, \citenamefont {Zalcman},
  \citenamefont {Neven},\ and\ \citenamefont {Martinis}}]{arute2019quantum}%
  \BibitemOpen
  \bibfield  {author} {\bibinfo {author} {\bibfnamefont {F.}~\bibnamefont
  {Arute}}, \bibinfo {author} {\bibfnamefont {K.}~\bibnamefont {Arya}},
  \bibinfo {author} {\bibfnamefont {R.}~\bibnamefont {Babbush}}, \bibinfo
  {author} {\bibfnamefont {D.}~\bibnamefont {Bacon}}, \bibinfo {author}
  {\bibfnamefont {J.~C.}\ \bibnamefont {Bardin}}, \bibinfo {author}
  {\bibfnamefont {R.}~\bibnamefont {Barends}}, \bibinfo {author} {\bibfnamefont
  {R.}~\bibnamefont {Biswas}}, \bibinfo {author} {\bibfnamefont
  {S.}~\bibnamefont {Boixo}}, \bibinfo {author} {\bibfnamefont {F.~G.}\
  \bibnamefont {Brandao}}, \bibinfo {author} {\bibfnamefont {D.~A.}\
  \bibnamefont {Buell}}, \bibinfo {author} {\bibfnamefont {B.}~\bibnamefont
  {Burkett}}, \bibinfo {author} {\bibfnamefont {Y.}~\bibnamefont {Chen}},
  \bibinfo {author} {\bibfnamefont {Z.}~\bibnamefont {Chen}}, \bibinfo {author}
  {\bibfnamefont {B.}~\bibnamefont {Chiaro}}, \bibinfo {author} {\bibfnamefont
  {R.}~\bibnamefont {Collins}}, \bibinfo {author} {\bibfnamefont
  {W.}~\bibnamefont {Courtney}}, \bibinfo {author} {\bibfnamefont
  {A.}~\bibnamefont {Dunsworth}}, \bibinfo {author} {\bibfnamefont
  {E.}~\bibnamefont {Farhi}}, \bibinfo {author} {\bibfnamefont
  {B.}~\bibnamefont {Foxen}}, \bibinfo {author} {\bibfnamefont
  {A.}~\bibnamefont {Fowler}}, \bibinfo {author} {\bibfnamefont
  {C.}~\bibnamefont {Gidney}}, \bibinfo {author} {\bibfnamefont
  {M.}~\bibnamefont {Giustina}}, \bibinfo {author} {\bibfnamefont
  {R.}~\bibnamefont {Graff}}, \bibinfo {author} {\bibfnamefont
  {K.}~\bibnamefont {Guerin}}, \bibinfo {author} {\bibfnamefont
  {S.}~\bibnamefont {Habegger}}, \bibinfo {author} {\bibfnamefont {M.~P.}\
  \bibnamefont {Harrigan}}, \bibinfo {author} {\bibfnamefont {M.~J.}\
  \bibnamefont {Hartmann}}, \bibinfo {author} {\bibfnamefont {A.}~\bibnamefont
  {Ho}}, \bibinfo {author} {\bibfnamefont {M.}~\bibnamefont {Hoffmann}},
  \bibinfo {author} {\bibfnamefont {T.}~\bibnamefont {Huang}}, \bibinfo
  {author} {\bibfnamefont {T.~S.}\ \bibnamefont {Humble}}, \bibinfo {author}
  {\bibfnamefont {S.~V.}\ \bibnamefont {Isakov}}, \bibinfo {author}
  {\bibfnamefont {E.}~\bibnamefont {Jeffrey}}, \bibinfo {author} {\bibfnamefont
  {Z.}~\bibnamefont {Jiang}}, \bibinfo {author} {\bibfnamefont
  {D.}~\bibnamefont {Kafri}}, \bibinfo {author} {\bibfnamefont
  {K.}~\bibnamefont {Kechedzhi}}, \bibinfo {author} {\bibfnamefont
  {J.}~\bibnamefont {Kelly}}, \bibinfo {author} {\bibfnamefont {P.~V.}\
  \bibnamefont {Klimov}}, \bibinfo {author} {\bibfnamefont {S.}~\bibnamefont
  {Knysh}}, \bibinfo {author} {\bibfnamefont {A.}~\bibnamefont {Korotkov}},
  \bibinfo {author} {\bibfnamefont {F.}~\bibnamefont {Kostritsa}}, \bibinfo
  {author} {\bibfnamefont {D.}~\bibnamefont {Landhuis}}, \bibinfo {author}
  {\bibfnamefont {M.}~\bibnamefont {Lindmark}}, \bibinfo {author}
  {\bibfnamefont {E.}~\bibnamefont {Lucero}}, \bibinfo {author} {\bibfnamefont
  {D.}~\bibnamefont {Lyakh}}, \bibinfo {author} {\bibfnamefont
  {S.}~\bibnamefont {Mandr¨¤}}, \bibinfo {author} {\bibfnamefont {J.~R.}\
  \bibnamefont {McClean}}, \bibinfo {author} {\bibfnamefont {M.}~\bibnamefont
  {McEwen}}, \bibinfo {author} {\bibfnamefont {A.}~\bibnamefont {Megrant}},
  \bibinfo {author} {\bibfnamefont {X.}~\bibnamefont {Mi}}, \bibinfo {author}
  {\bibfnamefont {K.}~\bibnamefont {Michielsen}}, \bibinfo {author}
  {\bibfnamefont {M.}~\bibnamefont {Mohseni}}, \bibinfo {author} {\bibfnamefont
  {J.}~\bibnamefont {Mutus}}, \bibinfo {author} {\bibfnamefont
  {O.}~\bibnamefont {Naaman}}, \bibinfo {author} {\bibfnamefont
  {M.}~\bibnamefont {Neeley}}, \bibinfo {author} {\bibfnamefont
  {C.}~\bibnamefont {Neill}}, \bibinfo {author} {\bibfnamefont {M.~Y.}\
  \bibnamefont {Niu}}, \bibinfo {author} {\bibfnamefont {E.}~\bibnamefont
  {Ostby}}, \bibinfo {author} {\bibfnamefont {A.}~\bibnamefont {Petukhov}},
  \bibinfo {author} {\bibfnamefont {J.~C.}\ \bibnamefont {Platt}}, \bibinfo
  {author} {\bibfnamefont {C.}~\bibnamefont {Quintana}}, \bibinfo {author}
  {\bibfnamefont {E.~G.}\ \bibnamefont {Rieffel}}, \bibinfo {author}
  {\bibfnamefont {P.}~\bibnamefont {Roushan}}, \bibinfo {author} {\bibfnamefont
  {N.~C.}\ \bibnamefont {Rubin}}, \bibinfo {author} {\bibfnamefont
  {D.}~\bibnamefont {Sank}}, \bibinfo {author} {\bibfnamefont {K.~J.}\
  \bibnamefont {Satzinger}}, \bibinfo {author} {\bibfnamefont {V.}~\bibnamefont
  {Smelyanskiy}}, \bibinfo {author} {\bibfnamefont {K.~J.}\ \bibnamefont
  {Sung}}, \bibinfo {author} {\bibfnamefont {M.~D.}\ \bibnamefont
  {Trevithick}}, \bibinfo {author} {\bibfnamefont {A.}~\bibnamefont
  {Vainsencher}}, \bibinfo {author} {\bibfnamefont {B.}~\bibnamefont
  {Villalonga}}, \bibinfo {author} {\bibfnamefont {T.}~\bibnamefont {White}},
  \bibinfo {author} {\bibfnamefont {Z.~J.}\ \bibnamefont {Yao}}, \bibinfo
  {author} {\bibfnamefont {P.}~\bibnamefont {Yeh}}, \bibinfo {author}
  {\bibfnamefont {A.}~\bibnamefont {Zalcman}}, \bibinfo {author} {\bibfnamefont
  {H.}~\bibnamefont {Neven}}, \ and\ \bibinfo {author} {\bibfnamefont {J.~M.}\
  \bibnamefont {Martinis}},\ }\href@noop {} {\bibfield  {journal} {\bibinfo
  {journal} {Nature}\ }\textbf {\bibinfo {volume} {574}},\ \bibinfo {pages}
  {505} (\bibinfo {year} {2019})}\BibitemShut {NoStop}%
\bibitem [{\citenamefont {Bennett}\ and\ \citenamefont
  {Brassard}(1984)}]{bennett1984quantum}%
  \BibitemOpen
  \bibfield  {author} {\bibinfo {author} {\bibfnamefont {C.~H.}\ \bibnamefont
  {Bennett}}\ and\ \bibinfo {author} {\bibfnamefont {G.}~\bibnamefont
  {Brassard}},\ }\href@noop {} {\bibfield  {journal} {\bibinfo  {journal}
  {Proc. Conf. Comp. Syst. Sig. Proc.}\ ,\ \bibinfo {pages} {175}} (\bibinfo
  {year} {IEEE, New York, 1984})}\BibitemShut {NoStop}%
\bibitem [{\citenamefont {Ekert}(1991)}]{ekert1991quantum}%
  \BibitemOpen
  \bibfield  {author} {\bibinfo {author} {\bibfnamefont {A.~K.}\ \bibnamefont
  {Ekert}},\ }\href@noop {} {\bibfield  {journal} {\bibinfo  {journal} {Phys.
  Rev. Lett.}\ }\textbf {\bibinfo {volume} {67}},\ \bibinfo {pages} {661}
  (\bibinfo {year} {1991})}\BibitemShut {NoStop}%
\bibitem [{\citenamefont {Sasaki}\ \emph {et~al.}(2014)\citenamefont {Sasaki},
  \citenamefont {Yamamoto},\ and\ \citenamefont
  {Koashi}}]{sasaki2014practical}%
  \BibitemOpen
  \bibfield  {author} {\bibinfo {author} {\bibfnamefont {T.}~\bibnamefont
  {Sasaki}}, \bibinfo {author} {\bibfnamefont {Y.}~\bibnamefont {Yamamoto}}, \
  and\ \bibinfo {author} {\bibfnamefont {M.}~\bibnamefont {Koashi}},\
  }\href@noop {} {\bibfield  {journal} {\bibinfo  {journal} {Nature}\ }\textbf
  {\bibinfo {volume} {509}},\ \bibinfo {pages} {475} (\bibinfo {year}
  {2014})}\BibitemShut {NoStop}%
\bibitem [{\citenamefont {Yin}\ \emph {et~al.}(2016{\natexlab{a}})\citenamefont
  {Yin}, \citenamefont {Fu}, \citenamefont {Mao},\ and\ \citenamefont
  {Chen}}]{yin2016detector}%
  \BibitemOpen
  \bibfield  {author} {\bibinfo {author} {\bibfnamefont {H.-L.}\ \bibnamefont
  {Yin}}, \bibinfo {author} {\bibfnamefont {Y.}~\bibnamefont {Fu}}, \bibinfo
  {author} {\bibfnamefont {Y.}~\bibnamefont {Mao}}, \ and\ \bibinfo {author}
  {\bibfnamefont {Z.-B.}\ \bibnamefont {Chen}},\ }\href@noop {} {\bibfield
  {journal} {\bibinfo  {journal} {Phys. Rev. A}\ }\textbf {\bibinfo {volume}
  {93}},\ \bibinfo {pages} {022330} (\bibinfo {year}
  {2016}{\natexlab{a}})}\BibitemShut {NoStop}%
\bibitem [{\citenamefont {Yin}\ \emph {et~al.}(2016{\natexlab{b}})\citenamefont
  {Yin}, \citenamefont {Chen}, \citenamefont {Yu}, \citenamefont {Liu},
  \citenamefont {You}, \citenamefont {Zhou}, \citenamefont {Chen},
  \citenamefont {Mao}, \citenamefont {Huang}, \citenamefont {Zhang},
  \citenamefont {Chen}, \citenamefont {Li}, \citenamefont {Nolan},
  \citenamefont {Zhou}, \citenamefont {Jiang}, \citenamefont {Wang},
  \citenamefont {Zhang}, \citenamefont {Wang},\ and\ \citenamefont
  {Pan}}]{yin2016measurement}%
  \BibitemOpen
  \bibfield  {author} {\bibinfo {author} {\bibfnamefont {H.-L.}\ \bibnamefont
  {Yin}}, \bibinfo {author} {\bibfnamefont {T.-Y.}\ \bibnamefont {Chen}},
  \bibinfo {author} {\bibfnamefont {Z.-W.}\ \bibnamefont {Yu}}, \bibinfo
  {author} {\bibfnamefont {H.}~\bibnamefont {Liu}}, \bibinfo {author}
  {\bibfnamefont {L.-X.}\ \bibnamefont {You}}, \bibinfo {author} {\bibfnamefont
  {Y.-H.}\ \bibnamefont {Zhou}}, \bibinfo {author} {\bibfnamefont {S.-J.}\
  \bibnamefont {Chen}}, \bibinfo {author} {\bibfnamefont {Y.}~\bibnamefont
  {Mao}}, \bibinfo {author} {\bibfnamefont {M.-Q.}\ \bibnamefont {Huang}},
  \bibinfo {author} {\bibfnamefont {W.-J.}\ \bibnamefont {Zhang}}, \bibinfo
  {author} {\bibfnamefont {H.}~\bibnamefont {Chen}}, \bibinfo {author}
  {\bibfnamefont {M.~J.}\ \bibnamefont {Li}}, \bibinfo {author} {\bibfnamefont
  {D.}~\bibnamefont {Nolan}}, \bibinfo {author} {\bibfnamefont
  {F.}~\bibnamefont {Zhou}}, \bibinfo {author} {\bibfnamefont {X.}~\bibnamefont
  {Jiang}}, \bibinfo {author} {\bibfnamefont {Z.}~\bibnamefont {Wang}},
  \bibinfo {author} {\bibfnamefont {Q.}~\bibnamefont {Zhang}}, \bibinfo
  {author} {\bibfnamefont {X.-B.}\ \bibnamefont {Wang}}, \ and\ \bibinfo
  {author} {\bibfnamefont {J.-W.}\ \bibnamefont {Pan}},\ }\href@noop {}
  {\bibfield  {journal} {\bibinfo  {journal} {Phys. Rev. Lett.}\ }\textbf
  {\bibinfo {volume} {117}},\ \bibinfo {pages} {190501} (\bibinfo {year}
  {2016}{\natexlab{b}})}\BibitemShut {NoStop}%
\bibitem [{\citenamefont {Boaron}\ \emph {et~al.}(2018)\citenamefont {Boaron},
  \citenamefont {Boso}, \citenamefont {Rusca}, \citenamefont {Vulliez},
  \citenamefont {Autebert}, \citenamefont {Caloz}, \citenamefont {Perrenoud},
  \citenamefont {Gras}, \citenamefont {Bussi\`eres}, \citenamefont {Li},
  \citenamefont {Nolan}, \citenamefont {Martin},\ and\ \citenamefont
  {Zbinden}}]{boaron2018secure}%
  \BibitemOpen
  \bibfield  {author} {\bibinfo {author} {\bibfnamefont {A.}~\bibnamefont
  {Boaron}}, \bibinfo {author} {\bibfnamefont {G.}~\bibnamefont {Boso}},
  \bibinfo {author} {\bibfnamefont {D.}~\bibnamefont {Rusca}}, \bibinfo
  {author} {\bibfnamefont {C.}~\bibnamefont {Vulliez}}, \bibinfo {author}
  {\bibfnamefont {C.}~\bibnamefont {Autebert}}, \bibinfo {author}
  {\bibfnamefont {M.}~\bibnamefont {Caloz}}, \bibinfo {author} {\bibfnamefont
  {M.}~\bibnamefont {Perrenoud}}, \bibinfo {author} {\bibfnamefont
  {G.}~\bibnamefont {Gras}}, \bibinfo {author} {\bibfnamefont {F.}~\bibnamefont
  {Bussi\`eres}}, \bibinfo {author} {\bibfnamefont {M.-J.}\ \bibnamefont {Li}},
  \bibinfo {author} {\bibfnamefont {D.}~\bibnamefont {Nolan}}, \bibinfo
  {author} {\bibfnamefont {A.}~\bibnamefont {Martin}}, \ and\ \bibinfo {author}
  {\bibfnamefont {H.}~\bibnamefont {Zbinden}},\ }\href@noop {} {\bibfield
  {journal} {\bibinfo  {journal} {Phys. Rev. Lett.}\ }\textbf {\bibinfo
  {volume} {121}},\ \bibinfo {pages} {190502} (\bibinfo {year}
  {2018})}\BibitemShut {NoStop}%
\bibitem [{\citenamefont {Fr{\"o}hlich}\ \emph {et~al.}(2013)\citenamefont
  {Fr{\"o}hlich}, \citenamefont {Dynes}, \citenamefont {Lucamarini},
  \citenamefont {Sharpe}, \citenamefont {Yuan},\ and\ \citenamefont
  {Shields}}]{frohlich2013quantum}%
  \BibitemOpen
  \bibfield  {author} {\bibinfo {author} {\bibfnamefont {B.}~\bibnamefont
  {Fr{\"o}hlich}}, \bibinfo {author} {\bibfnamefont {J.~F.}\ \bibnamefont
  {Dynes}}, \bibinfo {author} {\bibfnamefont {M.}~\bibnamefont {Lucamarini}},
  \bibinfo {author} {\bibfnamefont {A.~W.}\ \bibnamefont {Sharpe}}, \bibinfo
  {author} {\bibfnamefont {Z.}~\bibnamefont {Yuan}}, \ and\ \bibinfo {author}
  {\bibfnamefont {A.~J.}\ \bibnamefont {Shields}},\ }\href@noop {} {\bibfield
  {journal} {\bibinfo  {journal} {Nature}\ }\textbf {\bibinfo {volume} {501}},\
  \bibinfo {pages} {69} (\bibinfo {year} {2013})}\BibitemShut {NoStop}%
\bibitem [{\citenamefont {Liao}\ \emph {et~al.}(2018)\citenamefont {Liao},
  \citenamefont {Cai}, \citenamefont {Handsteiner}, \citenamefont {Liu},
  \citenamefont {Yin}, \citenamefont {Zhang}, \citenamefont {Rauch},
  \citenamefont {Fink}, \citenamefont {Ren}, \citenamefont {Liu}, \citenamefont
  {Li}, \citenamefont {Shen}, \citenamefont {Cao}, \citenamefont {Li},
  \citenamefont {Wang}, \citenamefont {Huang}, \citenamefont {Deng},
  \citenamefont {Xi}, \citenamefont {Ma}, \citenamefont {Hu}, \citenamefont
  {Li}, \citenamefont {Liu}, \citenamefont {Koidl}, \citenamefont {Wang},
  \citenamefont {Chen}, \citenamefont {Wang}, \citenamefont {Steindorfer},
  \citenamefont {Kirchner}, \citenamefont {Lu}, \citenamefont {Shu},
  \citenamefont {Ursin}, \citenamefont {Scheidl}, \citenamefont {Peng},
  \citenamefont {Wang}, \citenamefont {Zeilinger},\ and\ \citenamefont
  {Pan}}]{liao2018satellite}%
  \BibitemOpen
  \bibfield  {author} {\bibinfo {author} {\bibfnamefont {S.-K.}\ \bibnamefont
  {Liao}}, \bibinfo {author} {\bibfnamefont {W.-Q.}\ \bibnamefont {Cai}},
  \bibinfo {author} {\bibfnamefont {J.}~\bibnamefont {Handsteiner}}, \bibinfo
  {author} {\bibfnamefont {B.}~\bibnamefont {Liu}}, \bibinfo {author}
  {\bibfnamefont {J.}~\bibnamefont {Yin}}, \bibinfo {author} {\bibfnamefont
  {L.}~\bibnamefont {Zhang}}, \bibinfo {author} {\bibfnamefont
  {D.}~\bibnamefont {Rauch}}, \bibinfo {author} {\bibfnamefont
  {M.}~\bibnamefont {Fink}}, \bibinfo {author} {\bibfnamefont {J.-G.}\
  \bibnamefont {Ren}}, \bibinfo {author} {\bibfnamefont {W.-Y.}\ \bibnamefont
  {Liu}}, \bibinfo {author} {\bibfnamefont {Y.}~\bibnamefont {Li}}, \bibinfo
  {author} {\bibfnamefont {Q.}~\bibnamefont {Shen}}, \bibinfo {author}
  {\bibfnamefont {Y.}~\bibnamefont {Cao}}, \bibinfo {author} {\bibfnamefont
  {F.-Z.}\ \bibnamefont {Li}}, \bibinfo {author} {\bibfnamefont {J.-F.}\
  \bibnamefont {Wang}}, \bibinfo {author} {\bibfnamefont {Y.-M.}\ \bibnamefont
  {Huang}}, \bibinfo {author} {\bibfnamefont {L.}~\bibnamefont {Deng}},
  \bibinfo {author} {\bibfnamefont {T.}~\bibnamefont {Xi}}, \bibinfo {author}
  {\bibfnamefont {L.}~\bibnamefont {Ma}}, \bibinfo {author} {\bibfnamefont
  {T.}~\bibnamefont {Hu}}, \bibinfo {author} {\bibfnamefont {L.}~\bibnamefont
  {Li}}, \bibinfo {author} {\bibfnamefont {N.-L.}\ \bibnamefont {Liu}},
  \bibinfo {author} {\bibfnamefont {F.}~\bibnamefont {Koidl}}, \bibinfo
  {author} {\bibfnamefont {P.}~\bibnamefont {Wang}}, \bibinfo {author}
  {\bibfnamefont {Y.-A.}\ \bibnamefont {Chen}}, \bibinfo {author}
  {\bibfnamefont {X.-B.}\ \bibnamefont {Wang}}, \bibinfo {author}
  {\bibfnamefont {M.}~\bibnamefont {Steindorfer}}, \bibinfo {author}
  {\bibfnamefont {G.}~\bibnamefont {Kirchner}}, \bibinfo {author}
  {\bibfnamefont {C.-Y.}\ \bibnamefont {Lu}}, \bibinfo {author} {\bibfnamefont
  {R.}~\bibnamefont {Shu}}, \bibinfo {author} {\bibfnamefont {R.}~\bibnamefont
  {Ursin}}, \bibinfo {author} {\bibfnamefont {T.}~\bibnamefont {Scheidl}},
  \bibinfo {author} {\bibfnamefont {C.-Z.}\ \bibnamefont {Peng}}, \bibinfo
  {author} {\bibfnamefont {J.-Y.}\ \bibnamefont {Wang}}, \bibinfo {author}
  {\bibfnamefont {A.}~\bibnamefont {Zeilinger}}, \ and\ \bibinfo {author}
  {\bibfnamefont {J.-W.}\ \bibnamefont {Pan}},\ }\href@noop {} {\bibfield
  {journal} {\bibinfo  {journal} {Phys. Rev. Lett.}\ }\textbf {\bibinfo
  {volume} {120}},\ \bibinfo {pages} {030501} (\bibinfo {year}
  {2018})}\BibitemShut {NoStop}%
\bibitem [{\citenamefont {Yuan}\ \emph {et~al.}(2018)\citenamefont {Yuan},
  \citenamefont {Plews}, \citenamefont {Takahashi}, \citenamefont {Doi},
  \citenamefont {Tam}, \citenamefont {Sharpe}, \citenamefont {Dixon},
  \citenamefont {Lavelle}, \citenamefont {Dynes}, \citenamefont {Murakami},
  \citenamefont {Kujiraoka}, \citenamefont {Lucamarini}, \citenamefont
  {Tanizawa}, \citenamefont {Sato},\ and\ \citenamefont
  {Shields}}]{yuan201810}%
  \BibitemOpen
  \bibfield  {author} {\bibinfo {author} {\bibfnamefont {Z.-L.}\ \bibnamefont
  {Yuan}}, \bibinfo {author} {\bibfnamefont {A.}~\bibnamefont {Plews}},
  \bibinfo {author} {\bibfnamefont {R.}~\bibnamefont {Takahashi}}, \bibinfo
  {author} {\bibfnamefont {K.}~\bibnamefont {Doi}}, \bibinfo {author}
  {\bibfnamefont {W.}~\bibnamefont {Tam}}, \bibinfo {author} {\bibfnamefont
  {A.}~\bibnamefont {Sharpe}}, \bibinfo {author} {\bibfnamefont
  {A.}~\bibnamefont {Dixon}}, \bibinfo {author} {\bibfnamefont
  {E.}~\bibnamefont {Lavelle}}, \bibinfo {author} {\bibfnamefont
  {J.}~\bibnamefont {Dynes}}, \bibinfo {author} {\bibfnamefont
  {A.}~\bibnamefont {Murakami}}, \bibinfo {author} {\bibfnamefont
  {M.}~\bibnamefont {Kujiraoka}}, \bibinfo {author} {\bibfnamefont
  {M.}~\bibnamefont {Lucamarini}}, \bibinfo {author} {\bibfnamefont
  {Y.}~\bibnamefont {Tanizawa}}, \bibinfo {author} {\bibfnamefont
  {H.}~\bibnamefont {Sato}}, \ and\ \bibinfo {author} {\bibfnamefont {A.~J.}\
  \bibnamefont {Shields}},\ }\href@noop {} {\bibfield  {journal} {\bibinfo
  {journal} {J. Lightwave Technol.}\ }\textbf {\bibinfo {volume} {36}},\
  \bibinfo {pages} {3427} (\bibinfo {year} {2018})}\BibitemShut {NoStop}%
\bibitem [{\citenamefont {Zhang}\ \emph {et~al.}(2020)\citenamefont {Zhang},
  \citenamefont {Chen}, \citenamefont {Pirandola}, \citenamefont {Wang},
  \citenamefont {Zhou}, \citenamefont {Chu}, \citenamefont {Zhao},
  \citenamefont {Xu}, \citenamefont {Yu},\ and\ \citenamefont
  {Guo}}]{zhang2020long}%
  \BibitemOpen
  \bibfield  {author} {\bibinfo {author} {\bibfnamefont {Y.}~\bibnamefont
  {Zhang}}, \bibinfo {author} {\bibfnamefont {Z.}~\bibnamefont {Chen}},
  \bibinfo {author} {\bibfnamefont {S.}~\bibnamefont {Pirandola}}, \bibinfo
  {author} {\bibfnamefont {X.}~\bibnamefont {Wang}}, \bibinfo {author}
  {\bibfnamefont {C.}~\bibnamefont {Zhou}}, \bibinfo {author} {\bibfnamefont
  {B.}~\bibnamefont {Chu}}, \bibinfo {author} {\bibfnamefont {Y.}~\bibnamefont
  {Zhao}}, \bibinfo {author} {\bibfnamefont {B.}~\bibnamefont {Xu}}, \bibinfo
  {author} {\bibfnamefont {S.}~\bibnamefont {Yu}}, \ and\ \bibinfo {author}
  {\bibfnamefont {H.}~\bibnamefont {Guo}},\ }\href@noop {} {\bibfield
  {journal} {\bibinfo  {journal} {Phys. Rev. Lett.}\ }\textbf {\bibinfo
  {volume} {125}},\ \bibinfo {pages} {010502} (\bibinfo {year}
  {2020})}\BibitemShut {NoStop}%
\bibitem [{\citenamefont {Shamir}(1979)}]{shamir1979share}%
  \BibitemOpen
  \bibfield  {author} {\bibinfo {author} {\bibfnamefont {A.}~\bibnamefont
  {Shamir}},\ }\href@noop {} {\bibfield  {journal} {\bibinfo  {journal}
  {Commun. ACM}\ }\textbf {\bibinfo {volume} {22}},\ \bibinfo {pages} {612}
  (\bibinfo {year} {1979})}\BibitemShut {NoStop}%
\bibitem [{\citenamefont {Blakley}(1979)}]{blakley1979safeguarding}%
  \BibitemOpen
  \bibfield  {author} {\bibinfo {author} {\bibfnamefont {G.~R.}\ \bibnamefont
  {Blakley}},\ }\href@noop {} {\bibfield  {journal} {\bibinfo  {journal} {Proc.
  Natl Comp. Conf.}\ }\textbf {\bibinfo {volume} {48}},\ \bibinfo {pages} {313}
  (\bibinfo {year} {1979})}\BibitemShut {NoStop}%
\bibitem [{\citenamefont {Pirandola}\ \emph {et~al.}(2020)\citenamefont
  {Pirandola}, \citenamefont {Andersen}, \citenamefont {Banchi}, \citenamefont
  {Berta}, \citenamefont {Bunandar}, \citenamefont {Colbeck}, \citenamefont
  {Englund}, \citenamefont {Gehring}, \citenamefont {Lupo}, \citenamefont
  {Ottaviani}, \citenamefont {Pereira}, \citenamefont {Razavi}, \citenamefont
  {Shaari}, \citenamefont {Tomamichel}, \citenamefont {Usenko}, \citenamefont
  {Vallone}, \citenamefont {Villoresi},\ and\ \citenamefont
  {Wallden}}]{pirandola2020advances}%
  \BibitemOpen
  \bibfield  {author} {\bibinfo {author} {\bibfnamefont {S.}~\bibnamefont
  {Pirandola}}, \bibinfo {author} {\bibfnamefont {U.~L.}\ \bibnamefont
  {Andersen}}, \bibinfo {author} {\bibfnamefont {L.}~\bibnamefont {Banchi}},
  \bibinfo {author} {\bibfnamefont {M.}~\bibnamefont {Berta}}, \bibinfo
  {author} {\bibfnamefont {D.}~\bibnamefont {Bunandar}}, \bibinfo {author}
  {\bibfnamefont {R.}~\bibnamefont {Colbeck}}, \bibinfo {author} {\bibfnamefont
  {D.}~\bibnamefont {Englund}}, \bibinfo {author} {\bibfnamefont
  {T.}~\bibnamefont {Gehring}}, \bibinfo {author} {\bibfnamefont
  {C.}~\bibnamefont {Lupo}}, \bibinfo {author} {\bibfnamefont {C.}~\bibnamefont
  {Ottaviani}}, \bibinfo {author} {\bibfnamefont {J.~L.}\ \bibnamefont
  {Pereira}}, \bibinfo {author} {\bibfnamefont {M.}~\bibnamefont {Razavi}},
  \bibinfo {author} {\bibfnamefont {J.~S.}\ \bibnamefont {Shaari}}, \bibinfo
  {author} {\bibfnamefont {M.}~\bibnamefont {Tomamichel}}, \bibinfo {author}
  {\bibfnamefont {V.~C.}\ \bibnamefont {Usenko}}, \bibinfo {author}
  {\bibfnamefont {G.}~\bibnamefont {Vallone}}, \bibinfo {author} {\bibfnamefont
  {P.}~\bibnamefont {Villoresi}}, \ and\ \bibinfo {author} {\bibfnamefont
  {P.}~\bibnamefont {Wallden}},\ }\href@noop {} {\bibfield  {journal} {\bibinfo
   {journal} {Adv. Opt. Photon.}\ }\textbf {\bibinfo {volume} {12}},\ \bibinfo
  {pages} {1012} (\bibinfo {year} {2020})}\BibitemShut {NoStop}%
\bibitem [{\citenamefont {Hillery}\ \emph {et~al.}(1999)\citenamefont
  {Hillery}, \citenamefont {Bu{\v{z}}ek},\ and\ \citenamefont
  {Berthiaume}}]{hillery1999quantum}%
  \BibitemOpen
  \bibfield  {author} {\bibinfo {author} {\bibfnamefont {M.}~\bibnamefont
  {Hillery}}, \bibinfo {author} {\bibfnamefont {V.}~\bibnamefont
  {Bu{\v{z}}ek}}, \ and\ \bibinfo {author} {\bibfnamefont {A.}~\bibnamefont
  {Berthiaume}},\ }\href@noop {} {\bibfield  {journal} {\bibinfo  {journal}
  {Phys. Rev. A}\ }\textbf {\bibinfo {volume} {59}},\ \bibinfo {pages} {1829}
  (\bibinfo {year} {1999})}\BibitemShut {NoStop}%
\bibitem [{\citenamefont {Karlsson}\ \emph {et~al.}(1999)\citenamefont
  {Karlsson}, \citenamefont {Koashi},\ and\ \citenamefont
  {Imoto}}]{karlsson1999quantum}%
  \BibitemOpen
  \bibfield  {author} {\bibinfo {author} {\bibfnamefont {A.}~\bibnamefont
  {Karlsson}}, \bibinfo {author} {\bibfnamefont {M.}~\bibnamefont {Koashi}}, \
  and\ \bibinfo {author} {\bibfnamefont {N.}~\bibnamefont {Imoto}},\
  }\href@noop {} {\bibfield  {journal} {\bibinfo  {journal} {Phys. Rev. A}\
  }\textbf {\bibinfo {volume} {59}},\ \bibinfo {pages} {162} (\bibinfo {year}
  {1999})}\BibitemShut {NoStop}%
\bibitem [{\citenamefont {Chen}\ and\ \citenamefont
  {Lo}(2007)}]{chen2007multi}%
  \BibitemOpen
  \bibfield  {author} {\bibinfo {author} {\bibfnamefont {K.}~\bibnamefont
  {Chen}}\ and\ \bibinfo {author} {\bibfnamefont {H.-K.}\ \bibnamefont {Lo}},\
  }\href@noop {} {\bibfield  {journal} {\bibinfo  {journal} {Quantum Inf.
  Comput.}\ }\textbf {\bibinfo {volume} {7}},\ \bibinfo {pages} {689} (\bibinfo
  {year} {2007})}\BibitemShut {NoStop}%
\bibitem [{\citenamefont {Markham}\ and\ \citenamefont
  {Sanders}(2008)}]{markham2008graph}%
  \BibitemOpen
  \bibfield  {author} {\bibinfo {author} {\bibfnamefont {D.}~\bibnamefont
  {Markham}}\ and\ \bibinfo {author} {\bibfnamefont {B.~C.}\ \bibnamefont
  {Sanders}},\ }\href@noop {} {\bibfield  {journal} {\bibinfo  {journal} {Phys.
  Rev. A}\ }\textbf {\bibinfo {volume} {78}},\ \bibinfo {pages} {042309}
  (\bibinfo {year} {2008})}\BibitemShut {NoStop}%
\bibitem [{\citenamefont {Fu}\ \emph {et~al.}(2015)\citenamefont {Fu},
  \citenamefont {Yin}, \citenamefont {Chen},\ and\ \citenamefont
  {Chen}}]{fu2015long}%
  \BibitemOpen
  \bibfield  {author} {\bibinfo {author} {\bibfnamefont {Y.}~\bibnamefont
  {Fu}}, \bibinfo {author} {\bibfnamefont {H.-L.}\ \bibnamefont {Yin}},
  \bibinfo {author} {\bibfnamefont {T.-Y.}\ \bibnamefont {Chen}}, \ and\
  \bibinfo {author} {\bibfnamefont {Z.-B.}\ \bibnamefont {Chen}},\ }\href@noop
  {} {\bibfield  {journal} {\bibinfo  {journal} {Phys. Rev. Lett.}\ }\textbf
  {\bibinfo {volume} {114}},\ \bibinfo {pages} {090501} (\bibinfo {year}
  {2015})}\BibitemShut {NoStop}%
\bibitem [{\citenamefont {Tavakoli}\ \emph {et~al.}(2015)\citenamefont
  {Tavakoli}, \citenamefont {Herbauts}, \citenamefont {{\.Z}ukowski},\ and\
  \citenamefont {Bourennane}}]{tavakoli2015secret}%
  \BibitemOpen
  \bibfield  {author} {\bibinfo {author} {\bibfnamefont {A.}~\bibnamefont
  {Tavakoli}}, \bibinfo {author} {\bibfnamefont {I.}~\bibnamefont {Herbauts}},
  \bibinfo {author} {\bibfnamefont {M.}~\bibnamefont {{\.Z}ukowski}}, \ and\
  \bibinfo {author} {\bibfnamefont {M.}~\bibnamefont {Bourennane}},\
  }\href@noop {} {\bibfield  {journal} {\bibinfo  {journal} {Phys. Rev. A}\
  }\textbf {\bibinfo {volume} {92}},\ \bibinfo {pages} {030302} (\bibinfo
  {year} {2015})}\BibitemShut {NoStop}%
\bibitem [{\citenamefont {Kogias}\ \emph {et~al.}(2017)\citenamefont {Kogias},
  \citenamefont {Xiang}, \citenamefont {He},\ and\ \citenamefont
  {Adesso}}]{kogias2017unconditional}%
  \BibitemOpen
  \bibfield  {author} {\bibinfo {author} {\bibfnamefont {I.}~\bibnamefont
  {Kogias}}, \bibinfo {author} {\bibfnamefont {Y.}~\bibnamefont {Xiang}},
  \bibinfo {author} {\bibfnamefont {Q.}~\bibnamefont {He}}, \ and\ \bibinfo
  {author} {\bibfnamefont {G.}~\bibnamefont {Adesso}},\ }\href@noop {}
  {\bibfield  {journal} {\bibinfo  {journal} {Phys. Rev. A}\ }\textbf {\bibinfo
  {volume} {95}},\ \bibinfo {pages} {012315} (\bibinfo {year}
  {2017})}\BibitemShut {NoStop}%
\bibitem [{\citenamefont {Grice}\ and\ \citenamefont
  {Qi}(2019)}]{grice2019quantum}%
  \BibitemOpen
  \bibfield  {author} {\bibinfo {author} {\bibfnamefont {W.~P.}\ \bibnamefont
  {Grice}}\ and\ \bibinfo {author} {\bibfnamefont {B.}~\bibnamefont {Qi}},\
  }\href@noop {} {\bibfield  {journal} {\bibinfo  {journal} {Phys. Rev. A}\
  }\textbf {\bibinfo {volume} {100}},\ \bibinfo {pages} {022339} (\bibinfo
  {year} {2019})}\BibitemShut {NoStop}%
\bibitem [{\citenamefont {Wu}\ \emph {et~al.}(2020)\citenamefont {Wu},
  \citenamefont {Wang},\ and\ \citenamefont {Huang}}]{wu2020passive}%
  \BibitemOpen
  \bibfield  {author} {\bibinfo {author} {\bibfnamefont {X.}~\bibnamefont
  {Wu}}, \bibinfo {author} {\bibfnamefont {Y.}~\bibnamefont {Wang}}, \ and\
  \bibinfo {author} {\bibfnamefont {D.}~\bibnamefont {Huang}},\ }\href@noop {}
  {\bibfield  {journal} {\bibinfo  {journal} {Phy. Rev. A}\ }\textbf {\bibinfo
  {volume} {101}},\ \bibinfo {pages} {022301} (\bibinfo {year}
  {2020})}\BibitemShut {NoStop}%
\bibitem [{\citenamefont {Chen}\ \emph {et~al.}(2005)\citenamefont {Chen},
  \citenamefont {Zhang}, \citenamefont {Zhao}, \citenamefont {Zhou},
  \citenamefont {Lu}, \citenamefont {Peng}, \citenamefont {Yang},\ and\
  \citenamefont {Pan}}]{chen2005experimental}%
  \BibitemOpen
  \bibfield  {author} {\bibinfo {author} {\bibfnamefont {Y.-A.}\ \bibnamefont
  {Chen}}, \bibinfo {author} {\bibfnamefont {A.-N.}\ \bibnamefont {Zhang}},
  \bibinfo {author} {\bibfnamefont {Z.}~\bibnamefont {Zhao}}, \bibinfo {author}
  {\bibfnamefont {X.-Q.}\ \bibnamefont {Zhou}}, \bibinfo {author}
  {\bibfnamefont {C.-Y.}\ \bibnamefont {Lu}}, \bibinfo {author} {\bibfnamefont
  {C.-Z.}\ \bibnamefont {Peng}}, \bibinfo {author} {\bibfnamefont
  {T.}~\bibnamefont {Yang}}, \ and\ \bibinfo {author} {\bibfnamefont {J.-W.}\
  \bibnamefont {Pan}},\ }\href@noop {} {\bibfield  {journal} {\bibinfo
  {journal} {Phys. Rev. Lett.}\ }\textbf {\bibinfo {volume} {95}},\ \bibinfo
  {pages} {200502} (\bibinfo {year} {2005})}\BibitemShut {NoStop}%
\bibitem [{\citenamefont {Gaertner}\ \emph {et~al.}(2007)\citenamefont
  {Gaertner}, \citenamefont {Kurtsiefer}, \citenamefont {Bourennane},\ and\
  \citenamefont {Weinfurter}}]{gaertner2007experimental}%
  \BibitemOpen
  \bibfield  {author} {\bibinfo {author} {\bibfnamefont {S.}~\bibnamefont
  {Gaertner}}, \bibinfo {author} {\bibfnamefont {C.}~\bibnamefont
  {Kurtsiefer}}, \bibinfo {author} {\bibfnamefont {M.}~\bibnamefont
  {Bourennane}}, \ and\ \bibinfo {author} {\bibfnamefont {H.}~\bibnamefont
  {Weinfurter}},\ }\href@noop {} {\bibfield  {journal} {\bibinfo  {journal}
  {Phys. Rev. Lett.}\ }\textbf {\bibinfo {volume} {98}},\ \bibinfo {pages}
  {020503} (\bibinfo {year} {2007})}\BibitemShut {NoStop}%
\bibitem [{\citenamefont {Schmid}\ \emph {et~al.}(2005)\citenamefont {Schmid},
  \citenamefont {Trojek}, \citenamefont {Bourennane}, \citenamefont
  {Kurtsiefer}, \citenamefont {{\.Z}ukowski},\ and\ \citenamefont
  {Weinfurter}}]{schmid2005experimental}%
  \BibitemOpen
  \bibfield  {author} {\bibinfo {author} {\bibfnamefont {C.}~\bibnamefont
  {Schmid}}, \bibinfo {author} {\bibfnamefont {P.}~\bibnamefont {Trojek}},
  \bibinfo {author} {\bibfnamefont {M.}~\bibnamefont {Bourennane}}, \bibinfo
  {author} {\bibfnamefont {C.}~\bibnamefont {Kurtsiefer}}, \bibinfo {author}
  {\bibfnamefont {M.}~\bibnamefont {{\.Z}ukowski}}, \ and\ \bibinfo {author}
  {\bibfnamefont {H.}~\bibnamefont {Weinfurter}},\ }\href@noop {} {\bibfield
  {journal} {\bibinfo  {journal} {Phys. Rev. Lett.}\ }\textbf {\bibinfo
  {volume} {95}},\ \bibinfo {pages} {230505} (\bibinfo {year}
  {2005})}\BibitemShut {NoStop}%
\bibitem [{\citenamefont {Bell}\ \emph {et~al.}(2014)\citenamefont {Bell},
  \citenamefont {Markham}, \citenamefont {Herrera-Mart{\'\i}}, \citenamefont
  {Marin}, \citenamefont {Wadsworth}, \citenamefont {Rarity},\ and\
  \citenamefont {Tame}}]{bell2014experimental}%
  \BibitemOpen
  \bibfield  {author} {\bibinfo {author} {\bibfnamefont {B.}~\bibnamefont
  {Bell}}, \bibinfo {author} {\bibfnamefont {D.}~\bibnamefont {Markham}},
  \bibinfo {author} {\bibfnamefont {D.}~\bibnamefont {Herrera-Mart{\'\i}}},
  \bibinfo {author} {\bibfnamefont {A.}~\bibnamefont {Marin}}, \bibinfo
  {author} {\bibfnamefont {W.}~\bibnamefont {Wadsworth}}, \bibinfo {author}
  {\bibfnamefont {J.}~\bibnamefont {Rarity}}, \ and\ \bibinfo {author}
  {\bibfnamefont {M.}~\bibnamefont {Tame}},\ }\href@noop {} {\bibfield
  {journal} {\bibinfo  {journal} {Nat. Commun.}\ }\textbf {\bibinfo {volume}
  {5}},\ \bibinfo {pages} {5480} (\bibinfo {year} {2014})}\BibitemShut
  {NoStop}%
\bibitem [{\citenamefont {Cai}\ \emph {et~al.}(2017)\citenamefont {Cai},
  \citenamefont {Roslund}, \citenamefont {Ferrini}, \citenamefont {Arzani},
  \citenamefont {Xu}, \citenamefont {Fabre},\ and\ \citenamefont
  {Treps}}]{cai2017multimode}%
  \BibitemOpen
  \bibfield  {author} {\bibinfo {author} {\bibfnamefont {Y.}~\bibnamefont
  {Cai}}, \bibinfo {author} {\bibfnamefont {J.}~\bibnamefont {Roslund}},
  \bibinfo {author} {\bibfnamefont {G.}~\bibnamefont {Ferrini}}, \bibinfo
  {author} {\bibfnamefont {F.}~\bibnamefont {Arzani}}, \bibinfo {author}
  {\bibfnamefont {X.}~\bibnamefont {Xu}}, \bibinfo {author} {\bibfnamefont
  {C.}~\bibnamefont {Fabre}}, \ and\ \bibinfo {author} {\bibfnamefont
  {N.}~\bibnamefont {Treps}},\ }\href@noop {} {\bibfield  {journal} {\bibinfo
  {journal} {Nat. Commun.}\ }\textbf {\bibinfo {volume} {8}},\ \bibinfo {pages}
  {15645} (\bibinfo {year} {2017})}\BibitemShut {NoStop}%
\bibitem [{\citenamefont {Zhou}\ \emph {et~al.}(2018)\citenamefont {Zhou},
  \citenamefont {Yu}, \citenamefont {Yan}, \citenamefont {Jia}, \citenamefont
  {Zhang}, \citenamefont {Xie},\ and\ \citenamefont {Peng}}]{zhou2018quantum}%
  \BibitemOpen
  \bibfield  {author} {\bibinfo {author} {\bibfnamefont {Y.}~\bibnamefont
  {Zhou}}, \bibinfo {author} {\bibfnamefont {J.}~\bibnamefont {Yu}}, \bibinfo
  {author} {\bibfnamefont {Z.}~\bibnamefont {Yan}}, \bibinfo {author}
  {\bibfnamefont {X.}~\bibnamefont {Jia}}, \bibinfo {author} {\bibfnamefont
  {J.}~\bibnamefont {Zhang}}, \bibinfo {author} {\bibfnamefont
  {C.}~\bibnamefont {Xie}}, \ and\ \bibinfo {author} {\bibfnamefont
  {K.}~\bibnamefont {Peng}},\ }\href@noop {} {\bibfield  {journal} {\bibinfo
  {journal} {Phys. Rev. Lett.}\ }\textbf {\bibinfo {volume} {121}},\ \bibinfo
  {pages} {150502} (\bibinfo {year} {2018})}\BibitemShut {NoStop}%
\bibitem [{\citenamefont {Pan}\ \emph {et~al.}(2012)\citenamefont {Pan},
  \citenamefont {Chen}, \citenamefont {Lu}, \citenamefont {Weinfurter},
  \citenamefont {Zeilinger},\ and\ \citenamefont
  {{\.Z}ukowski}}]{pan2012multiphoton}%
  \BibitemOpen
  \bibfield  {author} {\bibinfo {author} {\bibfnamefont {J.-W.}\ \bibnamefont
  {Pan}}, \bibinfo {author} {\bibfnamefont {Z.-B.}\ \bibnamefont {Chen}},
  \bibinfo {author} {\bibfnamefont {C.-Y.}\ \bibnamefont {Lu}}, \bibinfo
  {author} {\bibfnamefont {H.}~\bibnamefont {Weinfurter}}, \bibinfo {author}
  {\bibfnamefont {A.}~\bibnamefont {Zeilinger}}, \ and\ \bibinfo {author}
  {\bibfnamefont {M.}~\bibnamefont {{\.Z}ukowski}},\ }\href@noop {} {\bibfield
  {journal} {\bibinfo  {journal} {Rev. Mod. Phys.}\ }\textbf {\bibinfo {volume}
  {84}},\ \bibinfo {pages} {777} (\bibinfo {year} {2012})}\BibitemShut
  {NoStop}%
\bibitem [{\citenamefont {Qin}\ \emph {et~al.}(2007)\citenamefont {Qin},
  \citenamefont {Gao}, \citenamefont {Wen},\ and\ \citenamefont
  {Zhu}}]{qin2007cryptanalysis}%
  \BibitemOpen
  \bibfield  {author} {\bibinfo {author} {\bibfnamefont {S.-J.}\ \bibnamefont
  {Qin}}, \bibinfo {author} {\bibfnamefont {F.}~\bibnamefont {Gao}}, \bibinfo
  {author} {\bibfnamefont {Q.-Y.}\ \bibnamefont {Wen}}, \ and\ \bibinfo
  {author} {\bibfnamefont {F.-C.}\ \bibnamefont {Zhu}},\ }\href@noop {}
  {\bibfield  {journal} {\bibinfo  {journal} {Phys. Rev. A}\ }\textbf {\bibinfo
  {volume} {76}},\ \bibinfo {pages} {062324} (\bibinfo {year}
  {2007})}\BibitemShut {NoStop}%
\bibitem [{\citenamefont {Wei}\ \emph {et~al.}(2018)\citenamefont {Wei},
  \citenamefont {Yang}, \citenamefont {Zhu},\ and\ \citenamefont
  {Yin}}]{wei2018quantum}%
  \BibitemOpen
  \bibfield  {author} {\bibinfo {author} {\bibfnamefont {K.-J.}\ \bibnamefont
  {Wei}}, \bibinfo {author} {\bibfnamefont {X.-Q.}\ \bibnamefont {Yang}},
  \bibinfo {author} {\bibfnamefont {C.-H.}\ \bibnamefont {Zhu}}, \ and\
  \bibinfo {author} {\bibfnamefont {Z.-Q.}\ \bibnamefont {Yin}},\ }\href@noop
  {} {\bibfield  {journal} {\bibinfo  {journal} {Quantum Inf. Process.}\
  }\textbf {\bibinfo {volume} {17}},\ \bibinfo {pages} {230} (\bibinfo {year}
  {2018})}\BibitemShut {NoStop}%
\bibitem [{\citenamefont {He}(2007)}]{he2007comment}%
  \BibitemOpen
  \bibfield  {author} {\bibinfo {author} {\bibfnamefont {G.~P.}\ \bibnamefont
  {He}},\ }\href@noop {} {\bibfield  {journal} {\bibinfo  {journal} {Phys. Rev.
  Lett.}\ }\textbf {\bibinfo {volume} {98}},\ \bibinfo {pages} {028901}
  (\bibinfo {year} {2007})}\BibitemShut {NoStop}%
\bibitem [{\citenamefont {Schmid}\ \emph {et~al.}(2007)\citenamefont {Schmid},
  \citenamefont {Trojek}, \citenamefont {Bourennane}, \citenamefont
  {Kurtsiefer}, \citenamefont {{\.Z}ukowski},\ and\ \citenamefont
  {Weinfurter}}]{schmid2007schmid}%
  \BibitemOpen
  \bibfield  {author} {\bibinfo {author} {\bibfnamefont {C.}~\bibnamefont
  {Schmid}}, \bibinfo {author} {\bibfnamefont {P.}~\bibnamefont {Trojek}},
  \bibinfo {author} {\bibfnamefont {M.}~\bibnamefont {Bourennane}}, \bibinfo
  {author} {\bibfnamefont {C.}~\bibnamefont {Kurtsiefer}}, \bibinfo {author}
  {\bibfnamefont {M.}~\bibnamefont {{\.Z}ukowski}}, \ and\ \bibinfo {author}
  {\bibfnamefont {H.}~\bibnamefont {Weinfurter}},\ }\href@noop {} {\bibfield
  {journal} {\bibinfo  {journal} {Phys. Rev. Lett.}\ }\textbf {\bibinfo
  {volume} {98}},\ \bibinfo {pages} {028902} (\bibinfo {year}
  {2007})}\BibitemShut {NoStop}%
\bibitem [{\citenamefont {Xu}\ \emph {et~al.}(2020)\citenamefont {Xu},
  \citenamefont {Ma}, \citenamefont {Zhang}, \citenamefont {Lo},\ and\
  \citenamefont {Pan}}]{xu2020secure}%
  \BibitemOpen
  \bibfield  {author} {\bibinfo {author} {\bibfnamefont {F.-H.}\ \bibnamefont
  {Xu}}, \bibinfo {author} {\bibfnamefont {X.-F.}\ \bibnamefont {Ma}}, \bibinfo
  {author} {\bibfnamefont {Q.}~\bibnamefont {Zhang}}, \bibinfo {author}
  {\bibfnamefont {H.-K.}\ \bibnamefont {Lo}}, \ and\ \bibinfo {author}
  {\bibfnamefont {J.-W.}\ \bibnamefont {Pan}},\ }\href@noop {} {\bibfield
  {journal} {\bibinfo  {journal} {Rev. Mod. Phys.}\ }\textbf {\bibinfo {volume}
  {92}},\ \bibinfo {pages} {025002} (\bibinfo {year} {2020})}\BibitemShut
  {NoStop}%
\bibitem [{\citenamefont {Lucamarini}\ \emph {et~al.}(2018)\citenamefont
  {Lucamarini}, \citenamefont {Yuan}, \citenamefont {Dynes},\ and\
  \citenamefont {Shields}}]{lucamarini2018overcoming}%
  \BibitemOpen
  \bibfield  {author} {\bibinfo {author} {\bibfnamefont {M.}~\bibnamefont
  {Lucamarini}}, \bibinfo {author} {\bibfnamefont {Z.~L.}\ \bibnamefont
  {Yuan}}, \bibinfo {author} {\bibfnamefont {J.~F.}\ \bibnamefont {Dynes}}, \
  and\ \bibinfo {author} {\bibfnamefont {A.~J.}\ \bibnamefont {Shields}},\
  }\href@noop {} {\bibfield  {journal} {\bibinfo  {journal} {Nature}\ }\textbf
  {\bibinfo {volume} {557}},\ \bibinfo {pages} {400} (\bibinfo {year}
  {2018})}\BibitemShut {NoStop}%
\bibitem [{\citenamefont {Wang}\ \emph {et~al.}(2018)\citenamefont {Wang},
  \citenamefont {Yu},\ and\ \citenamefont {Hu}}]{wang2018twin}%
  \BibitemOpen
  \bibfield  {author} {\bibinfo {author} {\bibfnamefont {X.-B.}\ \bibnamefont
  {Wang}}, \bibinfo {author} {\bibfnamefont {Z.-W.}\ \bibnamefont {Yu}}, \ and\
  \bibinfo {author} {\bibfnamefont {X.-L.}\ \bibnamefont {Hu}},\ }\href@noop {}
  {\bibfield  {journal} {\bibinfo  {journal} {Phys. Rev. A}\ }\textbf {\bibinfo
  {volume} {98}},\ \bibinfo {pages} {062323} (\bibinfo {year}
  {2018})}\BibitemShut {NoStop}%
\bibitem [{\citenamefont {Ma}\ \emph {et~al.}(2018)\citenamefont {Ma},
  \citenamefont {Zeng},\ and\ \citenamefont {Zhou}}]{ma2018phase}%
  \BibitemOpen
  \bibfield  {author} {\bibinfo {author} {\bibfnamefont {X.}~\bibnamefont
  {Ma}}, \bibinfo {author} {\bibfnamefont {P.}~\bibnamefont {Zeng}}, \ and\
  \bibinfo {author} {\bibfnamefont {H.}~\bibnamefont {Zhou}},\ }\href@noop {}
  {\bibfield  {journal} {\bibinfo  {journal} {Phys. Rev. X}\ }\textbf {\bibinfo
  {volume} {8}},\ \bibinfo {pages} {031043} (\bibinfo {year}
  {2018})}\BibitemShut {NoStop}%
\bibitem [{\citenamefont {Yin}\ and\ \citenamefont
  {Fu}(2019)}]{yin2019measurement}%
  \BibitemOpen
  \bibfield  {author} {\bibinfo {author} {\bibfnamefont {H.-L.}\ \bibnamefont
  {Yin}}\ and\ \bibinfo {author} {\bibfnamefont {Y.}~\bibnamefont {Fu}},\
  }\href@noop {} {\bibfield  {journal} {\bibinfo  {journal} {Sci. Rep.}\
  }\textbf {\bibinfo {volume} {9}},\ \bibinfo {pages} {3045} (\bibinfo {year}
  {2019})}\BibitemShut {NoStop}%
\bibitem [{\citenamefont {Lin}\ and\ \citenamefont
  {L{\"u}tkenhaus}(2018)}]{lin2018simple}%
  \BibitemOpen
  \bibfield  {author} {\bibinfo {author} {\bibfnamefont {J.}~\bibnamefont
  {Lin}}\ and\ \bibinfo {author} {\bibfnamefont {N.}~\bibnamefont
  {L{\"u}tkenhaus}},\ }\href@noop {} {\bibfield  {journal} {\bibinfo  {journal}
  {Phys. Rev. A}\ }\textbf {\bibinfo {volume} {98}},\ \bibinfo {pages} {042332}
  (\bibinfo {year} {2018})}\BibitemShut {NoStop}%
\bibitem [{\citenamefont {Cui}\ \emph {et~al.}(2019)\citenamefont {Cui},
  \citenamefont {Yin}, \citenamefont {Wang}, \citenamefont {Chen},
  \citenamefont {Wang}, \citenamefont {Guo},\ and\ \citenamefont
  {Han}}]{cui2019twin}%
  \BibitemOpen
  \bibfield  {author} {\bibinfo {author} {\bibfnamefont {C.}~\bibnamefont
  {Cui}}, \bibinfo {author} {\bibfnamefont {Z.-Q.}\ \bibnamefont {Yin}},
  \bibinfo {author} {\bibfnamefont {R.}~\bibnamefont {Wang}}, \bibinfo {author}
  {\bibfnamefont {W.}~\bibnamefont {Chen}}, \bibinfo {author} {\bibfnamefont
  {S.}~\bibnamefont {Wang}}, \bibinfo {author} {\bibfnamefont {G.-C.}\
  \bibnamefont {Guo}}, \ and\ \bibinfo {author} {\bibfnamefont {Z.-F.}\
  \bibnamefont {Han}},\ }\href@noop {} {\bibfield  {journal} {\bibinfo
  {journal} {Phys. Rev. Applied}\ }\textbf {\bibinfo {volume} {11}},\ \bibinfo
  {pages} {034053} (\bibinfo {year} {2019})}\BibitemShut {NoStop}%
\bibitem [{\citenamefont {Curty}\ \emph {et~al.}(2019)\citenamefont {Curty},
  \citenamefont {Azuma},\ and\ \citenamefont {Lo}}]{curty2019simple}%
  \BibitemOpen
  \bibfield  {author} {\bibinfo {author} {\bibfnamefont {M.}~\bibnamefont
  {Curty}}, \bibinfo {author} {\bibfnamefont {K.}~\bibnamefont {Azuma}}, \ and\
  \bibinfo {author} {\bibfnamefont {H.-K.}\ \bibnamefont {Lo}},\ }\href@noop {}
  {\bibfield  {journal} {\bibinfo  {journal} {npj Quantum Inf.}\ }\textbf
  {\bibinfo {volume} {5}},\ \bibinfo {pages} {64} (\bibinfo {year}
  {2019})}\BibitemShut {NoStop}%
\bibitem [{\citenamefont {Yin}\ and\ \citenamefont
  {Chen}(2019)}]{yin2019coherent}%
  \BibitemOpen
  \bibfield  {author} {\bibinfo {author} {\bibfnamefont {H.-L.}\ \bibnamefont
  {Yin}}\ and\ \bibinfo {author} {\bibfnamefont {Z.-B.}\ \bibnamefont {Chen}},\
  }\href@noop {} {\bibfield  {journal} {\bibinfo  {journal} {Sci. Rep.}\
  }\textbf {\bibinfo {volume} {9}},\ \bibinfo {pages} {14918} (\bibinfo {year}
  {2019})}\BibitemShut {NoStop}%
\bibitem [{\citenamefont {Maeda}\ \emph {et~al.}(2019)\citenamefont {Maeda},
  \citenamefont {Sasaki},\ and\ \citenamefont
  {Koashi}}]{maeda2019repeaterless}%
  \BibitemOpen
  \bibfield  {author} {\bibinfo {author} {\bibfnamefont {K.}~\bibnamefont
  {Maeda}}, \bibinfo {author} {\bibfnamefont {T.}~\bibnamefont {Sasaki}}, \
  and\ \bibinfo {author} {\bibfnamefont {M.}~\bibnamefont {Koashi}},\
  }\href@noop {} {\bibfield  {journal} {\bibinfo  {journal} {Nat. Commun.}\
  }\textbf {\bibinfo {volume} {10}},\ \bibinfo {pages} {3140} (\bibinfo {year}
  {2019})}\BibitemShut {NoStop}%
\bibitem [{\citenamefont {Pirandola}\ \emph {et~al.}(2017)\citenamefont
  {Pirandola}, \citenamefont {Laurenza}, \citenamefont {Ottaviani},\ and\
  \citenamefont {Banchi}}]{pirandola2017fundamental}%
  \BibitemOpen
  \bibfield  {author} {\bibinfo {author} {\bibfnamefont {S.}~\bibnamefont
  {Pirandola}}, \bibinfo {author} {\bibfnamefont {R.}~\bibnamefont {Laurenza}},
  \bibinfo {author} {\bibfnamefont {C.}~\bibnamefont {Ottaviani}}, \ and\
  \bibinfo {author} {\bibfnamefont {L.}~\bibnamefont {Banchi}},\ }\href@noop {}
  {\bibfield  {journal} {\bibinfo  {journal} {Nat. Commun.}\ }\textbf {\bibinfo
  {volume} {8}},\ \bibinfo {pages} {15043} (\bibinfo {year}
  {2017})}\BibitemShut {NoStop}%
\bibitem [{\citenamefont {Gu}\ \emph {et~al.}(2021)\citenamefont {Gu},
  \citenamefont {Cao}, \citenamefont {Yin},\ and\ \citenamefont
  {Chen}}]{Gu:21}%
  \BibitemOpen
  \bibfield  {author} {\bibinfo {author} {\bibfnamefont {J.}~\bibnamefont
  {Gu}}, \bibinfo {author} {\bibfnamefont {X.-Y.}\ \bibnamefont {Cao}},
  \bibinfo {author} {\bibfnamefont {H.-L.}\ \bibnamefont {Yin}}, \ and\
  \bibinfo {author} {\bibfnamefont {Z.-B.}\ \bibnamefont {Chen}},\ }\href
  {\doibase 10.1364/OE.417856} {\bibfield  {journal} {\bibinfo  {journal} {Opt.
  Express}\ }\textbf {\bibinfo {volume} {29}},\ \bibinfo {pages} {9165}
  (\bibinfo {year} {2021})}\BibitemShut {NoStop}%
\bibitem [{\citenamefont {Jia}\ \emph {et~al.}(2021)\citenamefont {Jia},
  \citenamefont {Gu}, \citenamefont {Li}, \citenamefont {Yin},\ and\
  \citenamefont {Chen}}]{e23060716}%
  \BibitemOpen
  \bibfield  {author} {\bibinfo {author} {\bibfnamefont {Z.-Y.}\ \bibnamefont
  {Jia}}, \bibinfo {author} {\bibfnamefont {J.}~\bibnamefont {Gu}}, \bibinfo
  {author} {\bibfnamefont {B.-H.}\ \bibnamefont {Li}}, \bibinfo {author}
  {\bibfnamefont {H.-L.}\ \bibnamefont {Yin}}, \ and\ \bibinfo {author}
  {\bibfnamefont {Z.-B.}\ \bibnamefont {Chen}},\ }\href@noop {} {\bibfield
  {journal} {\bibinfo  {journal} {Entropy}\ }\textbf {\bibinfo {volume} {23}},\
  \bibinfo {pages} {716} (\bibinfo {year} {2021})}\BibitemShut {NoStop}%
\bibitem [{\citenamefont {Takesue}\ \emph {et~al.}(2015)\citenamefont
  {Takesue}, \citenamefont {Sasaki}, \citenamefont {Tamaki},\ and\
  \citenamefont {Koashi}}]{takesue2015experimental}%
  \BibitemOpen
  \bibfield  {author} {\bibinfo {author} {\bibfnamefont {H.}~\bibnamefont
  {Takesue}}, \bibinfo {author} {\bibfnamefont {T.}~\bibnamefont {Sasaki}},
  \bibinfo {author} {\bibfnamefont {K.}~\bibnamefont {Tamaki}}, \ and\ \bibinfo
  {author} {\bibfnamefont {M.}~\bibnamefont {Koashi}},\ }\href@noop {}
  {\bibfield  {journal} {\bibinfo  {journal} {Nat. Photonics}\ }\textbf
  {\bibinfo {volume} {9}},\ \bibinfo {pages} {827} (\bibinfo {year}
  {2015})}\BibitemShut {NoStop}%
\bibitem [{\citenamefont {Wang}\ \emph {et~al.}(2015)\citenamefont {Wang},
  \citenamefont {Yin}, \citenamefont {Chen}, \citenamefont {He}, \citenamefont
  {Song}, \citenamefont {Li}, \citenamefont {Zhang}, \citenamefont {Zhou},
  \citenamefont {Guo},\ and\ \citenamefont {Han}}]{wang2015experimental}%
  \BibitemOpen
  \bibfield  {author} {\bibinfo {author} {\bibfnamefont {S.}~\bibnamefont
  {Wang}}, \bibinfo {author} {\bibfnamefont {Z.-Q.}\ \bibnamefont {Yin}},
  \bibinfo {author} {\bibfnamefont {W.}~\bibnamefont {Chen}}, \bibinfo {author}
  {\bibfnamefont {D.-Y.}\ \bibnamefont {He}}, \bibinfo {author} {\bibfnamefont
  {X.-T.}\ \bibnamefont {Song}}, \bibinfo {author} {\bibfnamefont {H.-W.}\
  \bibnamefont {Li}}, \bibinfo {author} {\bibfnamefont {L.-J.}\ \bibnamefont
  {Zhang}}, \bibinfo {author} {\bibfnamefont {Z.}~\bibnamefont {Zhou}},
  \bibinfo {author} {\bibfnamefont {G.-C.}\ \bibnamefont {Guo}}, \ and\
  \bibinfo {author} {\bibfnamefont {Z.-F.}\ \bibnamefont {Han}},\ }\href@noop
  {} {\bibfield  {journal} {\bibinfo  {journal} {Nat. Photonics}\ }\textbf
  {\bibinfo {volume} {9}},\ \bibinfo {pages} {832} (\bibinfo {year}
  {2015})}\BibitemShut {NoStop}%
\bibitem [{\citenamefont {Yin}\ \emph {et~al.}(2018)\citenamefont {Yin},
  \citenamefont {Wang}, \citenamefont {Chen}, \citenamefont {Han},
  \citenamefont {Wang}, \citenamefont {Guo},\ and\ \citenamefont
  {Han}}]{yin2018improved}%
  \BibitemOpen
  \bibfield  {author} {\bibinfo {author} {\bibfnamefont {Z.-Q.}\ \bibnamefont
  {Yin}}, \bibinfo {author} {\bibfnamefont {S.}~\bibnamefont {Wang}}, \bibinfo
  {author} {\bibfnamefont {W.}~\bibnamefont {Chen}}, \bibinfo {author}
  {\bibfnamefont {Y.-G.}\ \bibnamefont {Han}}, \bibinfo {author} {\bibfnamefont
  {R.}~\bibnamefont {Wang}}, \bibinfo {author} {\bibfnamefont {G.-C.}\
  \bibnamefont {Guo}}, \ and\ \bibinfo {author} {\bibfnamefont {Z.-F.}\
  \bibnamefont {Han}},\ }\href@noop {} {\bibfield  {journal} {\bibinfo
  {journal} {Nat. Commun.}\ }\textbf {\bibinfo {volume} {9}},\ \bibinfo {pages}
  {457} (\bibinfo {year} {2018})}\BibitemShut {NoStop}%
\bibitem [{\citenamefont {Minder}\ \emph {et~al.}(2019)\citenamefont {Minder},
  \citenamefont {Pittaluga}, \citenamefont {Roberts}, \citenamefont
  {Lucamarini}, \citenamefont {Dynes}, \citenamefont {Yuan},\ and\
  \citenamefont {Shields}}]{minder2019experimental}%
  \BibitemOpen
  \bibfield  {author} {\bibinfo {author} {\bibfnamefont {M.}~\bibnamefont
  {Minder}}, \bibinfo {author} {\bibfnamefont {M.}~\bibnamefont {Pittaluga}},
  \bibinfo {author} {\bibfnamefont {G.}~\bibnamefont {Roberts}}, \bibinfo
  {author} {\bibfnamefont {M.}~\bibnamefont {Lucamarini}}, \bibinfo {author}
  {\bibfnamefont {J.}~\bibnamefont {Dynes}}, \bibinfo {author} {\bibfnamefont
  {Z.}~\bibnamefont {Yuan}}, \ and\ \bibinfo {author} {\bibfnamefont
  {A.}~\bibnamefont {Shields}},\ }\href@noop {} {\bibfield  {journal} {\bibinfo
   {journal} {Nat. Photonics}\ }\textbf {\bibinfo {volume} {13}},\ \bibinfo
  {pages} {334} (\bibinfo {year} {2019})}\BibitemShut {NoStop}%
\bibitem [{\citenamefont {Zhong}\ \emph {et~al.}(2019)\citenamefont {Zhong},
  \citenamefont {Hu}, \citenamefont {Curty}, \citenamefont {Qian},\ and\
  \citenamefont {Lo}}]{zhong2019proof}%
  \BibitemOpen
  \bibfield  {author} {\bibinfo {author} {\bibfnamefont {X.}~\bibnamefont
  {Zhong}}, \bibinfo {author} {\bibfnamefont {J.}~\bibnamefont {Hu}}, \bibinfo
  {author} {\bibfnamefont {M.}~\bibnamefont {Curty}}, \bibinfo {author}
  {\bibfnamefont {L.}~\bibnamefont {Qian}}, \ and\ \bibinfo {author}
  {\bibfnamefont {H.-K.}\ \bibnamefont {Lo}},\ }\href@noop {} {\bibfield
  {journal} {\bibinfo  {journal} {Phys. Rev. Lett.}\ }\textbf {\bibinfo
  {volume} {123}},\ \bibinfo {pages} {100506} (\bibinfo {year}
  {2019})}\BibitemShut {NoStop}%
\bibitem [{\citenamefont {Wang}\ \emph {et~al.}(2019)\citenamefont {Wang},
  \citenamefont {He}, \citenamefont {Yin}, \citenamefont {Lu}, \citenamefont
  {Cui}, \citenamefont {Chen}, \citenamefont {Zhou}, \citenamefont {Guo},\ and\
  \citenamefont {Han}}]{wang2019beating}%
  \BibitemOpen
  \bibfield  {author} {\bibinfo {author} {\bibfnamefont {S.}~\bibnamefont
  {Wang}}, \bibinfo {author} {\bibfnamefont {D.-Y.}\ \bibnamefont {He}},
  \bibinfo {author} {\bibfnamefont {Z.-Q.}\ \bibnamefont {Yin}}, \bibinfo
  {author} {\bibfnamefont {F.-Y.}\ \bibnamefont {Lu}}, \bibinfo {author}
  {\bibfnamefont {C.-H.}\ \bibnamefont {Cui}}, \bibinfo {author} {\bibfnamefont
  {W.}~\bibnamefont {Chen}}, \bibinfo {author} {\bibfnamefont {Z.}~\bibnamefont
  {Zhou}}, \bibinfo {author} {\bibfnamefont {G.-C.}\ \bibnamefont {Guo}}, \
  and\ \bibinfo {author} {\bibfnamefont {Z.-F.}\ \bibnamefont {Han}},\
  }\href@noop {} {\bibfield  {journal} {\bibinfo  {journal} {Phys. Rev. X}\
  }\textbf {\bibinfo {volume} {9}},\ \bibinfo {pages} {021046} (\bibinfo {year}
  {2019})}\BibitemShut {NoStop}%
\bibitem [{\citenamefont {Chen}\ \emph {et~al.}(2020)\citenamefont {Chen},
  \citenamefont {Zhang}, \citenamefont {Liu}, \citenamefont {Jiang},
  \citenamefont {Zhang}, \citenamefont {Hu}, \citenamefont {Guan},
  \citenamefont {Yu}, \citenamefont {Xu}, \citenamefont {Lin}, \citenamefont
  {Li}, \citenamefont {Chen}, \citenamefont {Li}, \citenamefont {You},
  \citenamefont {Wang}, \citenamefont {Wang}, \citenamefont {Zhang},\ and\
  \citenamefont {Pan}}]{chen2020sending}%
  \BibitemOpen
  \bibfield  {author} {\bibinfo {author} {\bibfnamefont {J.-P.}\ \bibnamefont
  {Chen}}, \bibinfo {author} {\bibfnamefont {C.}~\bibnamefont {Zhang}},
  \bibinfo {author} {\bibfnamefont {Y.}~\bibnamefont {Liu}}, \bibinfo {author}
  {\bibfnamefont {C.}~\bibnamefont {Jiang}}, \bibinfo {author} {\bibfnamefont
  {W.}~\bibnamefont {Zhang}}, \bibinfo {author} {\bibfnamefont {X.-L.}\
  \bibnamefont {Hu}}, \bibinfo {author} {\bibfnamefont {J.-Y.}\ \bibnamefont
  {Guan}}, \bibinfo {author} {\bibfnamefont {Z.-W.}\ \bibnamefont {Yu}},
  \bibinfo {author} {\bibfnamefont {H.}~\bibnamefont {Xu}}, \bibinfo {author}
  {\bibfnamefont {J.}~\bibnamefont {Lin}}, \bibinfo {author} {\bibfnamefont
  {M.-J.}\ \bibnamefont {Li}}, \bibinfo {author} {\bibfnamefont
  {H.}~\bibnamefont {Chen}}, \bibinfo {author} {\bibfnamefont {H.}~\bibnamefont
  {Li}}, \bibinfo {author} {\bibfnamefont {L.}~\bibnamefont {You}}, \bibinfo
  {author} {\bibfnamefont {Z.}~\bibnamefont {Wang}}, \bibinfo {author}
  {\bibfnamefont {X.-B.}\ \bibnamefont {Wang}}, \bibinfo {author}
  {\bibfnamefont {Q.}~\bibnamefont {Zhang}}, \ and\ \bibinfo {author}
  {\bibfnamefont {J.-W.}\ \bibnamefont {Pan}},\ }\href@noop {} {\bibfield
  {journal} {\bibinfo  {journal} {Phys. Rev. Lett.}\ }\textbf {\bibinfo
  {volume} {124}},\ \bibinfo {pages} {070501} (\bibinfo {year}
  {2020})}\BibitemShut {NoStop}%
\bibitem [{\citenamefont {Fang}\ \emph {et~al.}(2020)\citenamefont {Fang},
  \citenamefont {Zeng}, \citenamefont {Liu}, \citenamefont {Zou}, \citenamefont
  {Wu}, \citenamefont {Tang}, \citenamefont {Sheng}, \citenamefont {Xiang},
  \citenamefont {Zhang}, \citenamefont {Li}, \citenamefont {Wang},
  \citenamefont {You}, \citenamefont {Li}, \citenamefont {Chen}, \citenamefont
  {Chen}, \citenamefont {Zhang}, \citenamefont {Peng}, \citenamefont {Ma},
  \citenamefont {Chen},\ and\ \citenamefont {Pan}}]{fang2020implementation}%
  \BibitemOpen
  \bibfield  {author} {\bibinfo {author} {\bibfnamefont {X.-T.}\ \bibnamefont
  {Fang}}, \bibinfo {author} {\bibfnamefont {P.}~\bibnamefont {Zeng}}, \bibinfo
  {author} {\bibfnamefont {H.}~\bibnamefont {Liu}}, \bibinfo {author}
  {\bibfnamefont {M.}~\bibnamefont {Zou}}, \bibinfo {author} {\bibfnamefont
  {W.-J.}\ \bibnamefont {Wu}}, \bibinfo {author} {\bibfnamefont {Y.-L.}\
  \bibnamefont {Tang}}, \bibinfo {author} {\bibfnamefont {Y.-J.}\ \bibnamefont
  {Sheng}}, \bibinfo {author} {\bibfnamefont {Y.}~\bibnamefont {Xiang}},
  \bibinfo {author} {\bibfnamefont {W.}~\bibnamefont {Zhang}}, \bibinfo
  {author} {\bibfnamefont {H.}~\bibnamefont {Li}}, \bibinfo {author}
  {\bibfnamefont {Z.}~\bibnamefont {Wang}}, \bibinfo {author} {\bibfnamefont
  {L.}~\bibnamefont {You}}, \bibinfo {author} {\bibfnamefont {M.-J.}\
  \bibnamefont {Li}}, \bibinfo {author} {\bibfnamefont {H.}~\bibnamefont
  {Chen}}, \bibinfo {author} {\bibfnamefont {Y.-A.}\ \bibnamefont {Chen}},
  \bibinfo {author} {\bibfnamefont {Q.}~\bibnamefont {Zhang}}, \bibinfo
  {author} {\bibfnamefont {C.-Z.}\ \bibnamefont {Peng}}, \bibinfo {author}
  {\bibfnamefont {X.}~\bibnamefont {Ma}}, \bibinfo {author} {\bibfnamefont
  {T.-Y.}\ \bibnamefont {Chen}}, \ and\ \bibinfo {author} {\bibfnamefont
  {J.-W.}\ \bibnamefont {Pan}},\ }\href@noop {} {\bibfield  {journal} {\bibinfo
   {journal} {Nat. Photonics}\ }\textbf {\bibinfo {volume} {14}},\ \bibinfo
  {pages} {422} (\bibinfo {year} {2020})}\BibitemShut {NoStop}%
\bibitem [{\citenamefont {Arrazola}\ and\ \citenamefont
  {L\"utkenhaus}(2014)}]{Arrazola2014Quantum}%
  \BibitemOpen
  \bibfield  {author} {\bibinfo {author} {\bibfnamefont {J.~M.}\ \bibnamefont
  {Arrazola}}\ and\ \bibinfo {author} {\bibfnamefont {N.}~\bibnamefont
  {L\"utkenhaus}},\ }\href@noop {} {\bibfield  {journal} {\bibinfo  {journal}
  {Phys. Rev. A}\ }\textbf {\bibinfo {volume} {89}},\ \bibinfo {pages} {062305}
  (\bibinfo {year} {2014})}\BibitemShut {NoStop}%
\bibitem [{\citenamefont {Guan}\ \emph {et~al.}(2016)\citenamefont {Guan},
  \citenamefont {Xu}, \citenamefont {Yin}, \citenamefont {Li}, \citenamefont
  {Zhang}, \citenamefont {Chen}, \citenamefont {Yang}, \citenamefont {Li},
  \citenamefont {You}, \citenamefont {Chen}, \citenamefont {Wang},
  \citenamefont {Zhang},\ and\ \citenamefont {Pan}}]{Guan2016Observation}%
  \BibitemOpen
  \bibfield  {author} {\bibinfo {author} {\bibfnamefont {J.-Y.}\ \bibnamefont
  {Guan}}, \bibinfo {author} {\bibfnamefont {F.}~\bibnamefont {Xu}}, \bibinfo
  {author} {\bibfnamefont {H.-L.}\ \bibnamefont {Yin}}, \bibinfo {author}
  {\bibfnamefont {Y.}~\bibnamefont {Li}}, \bibinfo {author} {\bibfnamefont
  {W.-J.}\ \bibnamefont {Zhang}}, \bibinfo {author} {\bibfnamefont {S.-J.}\
  \bibnamefont {Chen}}, \bibinfo {author} {\bibfnamefont {X.-Y.}\ \bibnamefont
  {Yang}}, \bibinfo {author} {\bibfnamefont {L.}~\bibnamefont {Li}}, \bibinfo
  {author} {\bibfnamefont {L.-X.}\ \bibnamefont {You}}, \bibinfo {author}
  {\bibfnamefont {T.-Y.}\ \bibnamefont {Chen}}, \bibinfo {author}
  {\bibfnamefont {Z.}~\bibnamefont {Wang}}, \bibinfo {author} {\bibfnamefont
  {Q.}~\bibnamefont {Zhang}}, \ and\ \bibinfo {author} {\bibfnamefont {J.-W.}\
  \bibnamefont {Pan}},\ }\href@noop {} {\bibfield  {journal} {\bibinfo
  {journal} {Phys. Rev. Lett.}\ }\textbf {\bibinfo {volume} {116}},\ \bibinfo
  {pages} {240502} (\bibinfo {year} {2016})}\BibitemShut {NoStop}%
\bibitem [{\citenamefont {Koashi}(2009)}]{Koashi2009simple}%
  \BibitemOpen
  \bibfield  {author} {\bibinfo {author} {\bibfnamefont {M.}~\bibnamefont
  {Koashi}},\ }\href@noop {} {\bibfield  {journal} {\bibinfo  {journal} {New J.
  Phys.}\ }\textbf {\bibinfo {volume} {11}},\ \bibinfo {pages} {045018}
  (\bibinfo {year} {2009})}\BibitemShut {NoStop}%
\bibitem [{\citenamefont {Takeoka}\ \emph {et~al.}(2014)\citenamefont
  {Takeoka}, \citenamefont {Guha},\ and\ \citenamefont
  {Wilde}}]{takeoka2014fundamental}%
  \BibitemOpen
  \bibfield  {author} {\bibinfo {author} {\bibfnamefont {M.}~\bibnamefont
  {Takeoka}}, \bibinfo {author} {\bibfnamefont {S.}~\bibnamefont {Guha}}, \
  and\ \bibinfo {author} {\bibfnamefont {M.~M.}\ \bibnamefont {Wilde}},\
  }\href@noop {} {\bibfield  {journal} {\bibinfo  {journal} {Nat. Commun.}\
  }\textbf {\bibinfo {volume} {5}},\ \bibinfo {pages} {5235} (\bibinfo {year}
  {2014})}\BibitemShut {NoStop}%
\bibitem [{\citenamefont {Bouchard}\ \emph {et~al.}(2018)\citenamefont
  {Bouchard}, \citenamefont {Sit}, \citenamefont {Heshami}, \citenamefont
  {Fickler},\ and\ \citenamefont {Karimi}}]{bouchard2018round}%
  \BibitemOpen
  \bibfield  {author} {\bibinfo {author} {\bibfnamefont {F.}~\bibnamefont
  {Bouchard}}, \bibinfo {author} {\bibfnamefont {A.}~\bibnamefont {Sit}},
  \bibinfo {author} {\bibfnamefont {K.}~\bibnamefont {Heshami}}, \bibinfo
  {author} {\bibfnamefont {R.}~\bibnamefont {Fickler}}, \ and\ \bibinfo
  {author} {\bibfnamefont {E.}~\bibnamefont {Karimi}},\ }\href@noop {}
  {\bibfield  {journal} {\bibinfo  {journal} {Phys. Rev. A}\ }\textbf {\bibinfo
  {volume} {98}},\ \bibinfo {pages} {010301} (\bibinfo {year}
  {2018})}\BibitemShut {NoStop}%
\bibitem [{\citenamefont {Cleve}\ \emph {et~al.}(1999)\citenamefont {Cleve},
  \citenamefont {Gottesman},\ and\ \citenamefont {Lo}}]{Cleve:1999:How}%
  \BibitemOpen
  \bibfield  {author} {\bibinfo {author} {\bibfnamefont {R.}~\bibnamefont
  {Cleve}}, \bibinfo {author} {\bibfnamefont {D.}~\bibnamefont {Gottesman}}, \
  and\ \bibinfo {author} {\bibfnamefont {H.-K.}\ \bibnamefont {Lo}},\
  }\href@noop {} {\bibfield  {journal} {\bibinfo  {journal} {Phys. Rev. Lett.}\
  }\textbf {\bibinfo {volume} {83}},\ \bibinfo {pages} {648} (\bibinfo {year}
  {1999})}\BibitemShut {NoStop}%
\bibitem [{\citenamefont {{\.Z}ukowski}\ \emph {et~al.}(1998)\citenamefont
  {{\.Z}ukowski}, \citenamefont {Zeilinger}, \citenamefont {Horne},\ and\
  \citenamefont {Weinfurter}}]{zukowski1998quest}%
  \BibitemOpen
  \bibfield  {author} {\bibinfo {author} {\bibfnamefont {M.}~\bibnamefont
  {{\.Z}ukowski}}, \bibinfo {author} {\bibfnamefont {A.}~\bibnamefont
  {Zeilinger}}, \bibinfo {author} {\bibfnamefont {M.}~\bibnamefont {Horne}}, \
  and\ \bibinfo {author} {\bibfnamefont {H.}~\bibnamefont {Weinfurter}},\
  }\href@noop {} {\bibfield  {journal} {\bibinfo  {journal} {Acta Phys. Pol.
  A}\ }\textbf {\bibinfo {volume} {93}},\ \bibinfo {pages} {187} (\bibinfo
  {year} {1998})}\BibitemShut {NoStop}%
\bibitem [{\citenamefont {Chen}\ \emph {et~al.}(2021)\citenamefont {Chen},
  \citenamefont {Zhang}, \citenamefont {Chen}, \citenamefont {Cai},
  \citenamefont {Liao}, \citenamefont {Zhang}, \citenamefont {Chen},
  \citenamefont {Yin}, \citenamefont {Ren}, \citenamefont {Chen}, \citenamefont
  {Han}, \citenamefont {Yu}, \citenamefont {Liang}, \citenamefont {Zhou},
  \citenamefont {Yuan}, \citenamefont {Zhao}, \citenamefont {Wang},
  \citenamefont {Jiang}, \citenamefont {Zhang}, \citenamefont {Liu},
  \citenamefont {Li}, \citenamefont {Shen}, \citenamefont {Cao}, \citenamefont
  {Lu}, \citenamefont {Shu}, \citenamefont {Wang}, \citenamefont {Li},
  \citenamefont {Liu}, \citenamefont {Xu}, \citenamefont {Wang}, \citenamefont
  {Peng},\ and\ \citenamefont {Pan}}]{chen2021integrated}%
  \BibitemOpen
  \bibfield  {author} {\bibinfo {author} {\bibfnamefont {Y.-A.}\ \bibnamefont
  {Chen}}, \bibinfo {author} {\bibfnamefont {Q.}~\bibnamefont {Zhang}},
  \bibinfo {author} {\bibfnamefont {T.-Y.}\ \bibnamefont {Chen}}, \bibinfo
  {author} {\bibfnamefont {W.-Q.}\ \bibnamefont {Cai}}, \bibinfo {author}
  {\bibfnamefont {S.-K.}\ \bibnamefont {Liao}}, \bibinfo {author}
  {\bibfnamefont {J.}~\bibnamefont {Zhang}}, \bibinfo {author} {\bibfnamefont
  {K.}~\bibnamefont {Chen}}, \bibinfo {author} {\bibfnamefont {J.}~\bibnamefont
  {Yin}}, \bibinfo {author} {\bibfnamefont {J.-G.}\ \bibnamefont {Ren}},
  \bibinfo {author} {\bibfnamefont {Z.}~\bibnamefont {Chen}}, \bibinfo {author}
  {\bibfnamefont {S.-L.}\ \bibnamefont {Han}}, \bibinfo {author} {\bibfnamefont
  {Q.}~\bibnamefont {Yu}}, \bibinfo {author} {\bibfnamefont {K.}~\bibnamefont
  {Liang}}, \bibinfo {author} {\bibfnamefont {F.}~\bibnamefont {Zhou}},
  \bibinfo {author} {\bibfnamefont {X.}~\bibnamefont {Yuan}}, \bibinfo {author}
  {\bibfnamefont {M.-S.}\ \bibnamefont {Zhao}}, \bibinfo {author}
  {\bibfnamefont {T.-Y.}\ \bibnamefont {Wang}}, \bibinfo {author}
  {\bibfnamefont {X.}~\bibnamefont {Jiang}}, \bibinfo {author} {\bibfnamefont
  {L.}~\bibnamefont {Zhang}}, \bibinfo {author} {\bibfnamefont {W.-Y.}\
  \bibnamefont {Liu}}, \bibinfo {author} {\bibfnamefont {Y.}~\bibnamefont
  {Li}}, \bibinfo {author} {\bibfnamefont {Q.}~\bibnamefont {Shen}}, \bibinfo
  {author} {\bibfnamefont {Y.}~\bibnamefont {Cao}}, \bibinfo {author}
  {\bibfnamefont {C.-Y.}\ \bibnamefont {Lu}}, \bibinfo {author} {\bibfnamefont
  {R.}~\bibnamefont {Shu}}, \bibinfo {author} {\bibfnamefont {J.-Y.}\
  \bibnamefont {Wang}}, \bibinfo {author} {\bibfnamefont {L.}~\bibnamefont
  {Li}}, \bibinfo {author} {\bibfnamefont {N.-L.}\ \bibnamefont {Liu}},
  \bibinfo {author} {\bibfnamefont {F.}~\bibnamefont {Xu}}, \bibinfo {author}
  {\bibfnamefont {X.-B.}\ \bibnamefont {Wang}}, \bibinfo {author}
  {\bibfnamefont {C.-Z.}\ \bibnamefont {Peng}}, \ and\ \bibinfo {author}
  {\bibfnamefont {J.-W.}\ \bibnamefont {Pan}},\ }\href@noop {} {\bibfield
  {journal} {\bibinfo  {journal} {Nature}\ }\textbf {\bibinfo {volume} {589}},\
  \bibinfo {pages} {214} (\bibinfo {year} {2021})}\BibitemShut {NoStop}%
\end{thebibliography}

%

\end{document}